\documentclass[a4paper,11pt]{article}

\usepackage{amsmath,amssymb,amsfonts,amsbsy,amsthm,booktabs,epsfig,graphicx,rotating,microtype}
\usepackage[textwidth=34pc,textheight=650pt]{geometry}
\usepackage{authblk}

\input{util/commands.tex}

\usepackage[backend=biber,      
	bibstyle=ieee,              
	citestyle=numeric-comp,     
	sortcites=true,
	maxbibnames=3,
	bibencoding=utf8]{biblatex}
\begin{document}

\title{A Continuous Benchmarking Infrastructure for High-Performance Computing Applications}

\author[a,b,$\ast$]{Christoph Alt\orcidlink{0000-0001-8897-5205}}
\author[c,d]{Martin Lanser\orcidlink{0000-0002-4232-9395}}
\author[a]{Jonas Plewinski\orcidlink{0000-0002-4815-7643}}
\author[f]{Atin Janki\orcidlink{0009-0003-0429-9804}}
\author[c,d]{Axel Klawonn\orcidlink{0000-0003-4765-7387}}
\author[a,e]{Harald K\"ostler\orcidlink{0000-0002-6992-2690}}
\author[g]{Michael Selzer\orcidlink{0000-0002-9756-646X}}
\author[a,h]{Ulrich R{\"u}de\orcidlink{0000-0001-8796-8599}}
\affil[a]{Chair for Computer Science 10 (System Simulation), Friedrich-Alexander-Universität Erlangen-Nürnberg, Cauerstraße 11, 91058 Erlangen, Germany}
\affil[b]{Paderborn University, Paderborn Center for Parallel Computing, Warburger Straße 100, 33098 Paderborn, Germany}
\affil[c]{Department of Mathematics and Computer Science, University of Cologne, Weyertal 86-90, 50931 K\"oln, Germany}
\affil[d]{Center for Data and Simulation Science, University of Cologne, Albertus-Magnus-Platz, 50923 K\"oln, Germany}
\affil[e]{Erlangen National High Performance Computing Center (NHR@FAU), Martensstraße 1, 91058 Erlangen, Germany}
\affil[f]{Institute for Applied Materials (IAM), Karlsruhe Institute of Technology (KIT), Straße am Forum 7, 76131 Karlsruhe, Germany}
\affil[g]{Institute of Nanotechnology (INT), Karlsruhe Institute of Technology (KIT), Hermann-von-Helmholtz-Platz 1, 76344 Eggenstein-Leopoldshafen, Germany}
\affil[h]{CERFACS, 42 Avenue Gaspard Coriolis, 31057 Toulouse Cedex 1, France}
\affil[$\ast$]{CONTACT C. Alt. Email: \texttt{christoph.alt@fau.de}}

\maketitle

\begin{abstract}
	For scientific software, especially those used for large-scale simulations, achieving good performance and efficiently using the available hardware resources is essential.
	It is important to regularly perform benchmarks to ensure the efficient use of hardware and software when systems are changing and the software evolves.
	However, this can become quickly very tedious when many options for parameters, solvers, and hardware architectures are available.
	We present a continuous benchmarking strategy that automates benchmarking new code changes on high-performance computing clusters.
	This makes it possible to track how each code change affects the performance and how it evolves.
\end{abstract}


\acresetall{}

\section{Introduction}

Scientific software, specifically large-scale simulations, often require enormous compute power so that they can only be run on advanced parallel supercomputers.
This incurs a high resource consumption, both in terms of proportionate infrastructure expenses as well as operating costs.
In particular the energy bill for operating high-performance computers is becoming a critical factor, both financially and environmentally.

When a computational campaign, say, e.g., a parameter study for an engineering design, consumes supercomputer resources worth a million of Dollars or Euros, the natural question would be:
could the same result be achieved more cheaply with an optimized code or maybe a different, better designed code?
This obvious question, however, seems to be rarely asked.
Nevertheless, a 10\% improvement would yield substantial savings for such a compute-intensive task.
Note also that the loss may not only be financial, but that also the quality of scientific results may be compromised when, e.g., the necessary resolution of a physical model cannot be reached due to a poor code performance.
It is unsatisfactory that such cases may exist, but in many cases neither the users nor the developers of the code have a way to know since no systematic analysis is available.

In our experience, even widely used scientific codes can often be accelerated significantly through systematic architecture aware optimization.
The techniques of code optimization may be as trivial as choosing compiler options appropriate for a given hardware or by selecting better algorithmic options, e.g., choosing more efficient linear solvers.
Possible code optimization may involve classical techniques such as exploiting vectorization and removing parallel load imbalances.
Further steps may involve highly complex modifications to a program, such as changing the underlying data structures to make it more appropriate for a given hardware.
In its most extreme form, an optimization could result in a complete redesign of a code.

Even once a code is optimized, there is no guarantee that the performance will remain optimal throughout development.
Future enhancements to the software, including support for new simulation scenarios or new hardware architectures, can - often unintentionally - result in a performance loss.
It is necessary to link the process of systematic performance analysis closely to the development process.
In practice, testing many combinations of different parameters on various hardware options can become tedious.

This article addresses the performance gap in scientific computing codes by proposing a {\em \acl{cb}} strategy.
Here, we understand \emph{\acl{cb}} as analogous to \emph{\acl{ci}}, primarily a paradigm to check a code's functional properties and correctness as integral steps of the development process.
{\em \Acl{cb}} extends this paradigm to systematically evaluate and analyse the performance properties of a code as a means to give developers and users insight and early feedback when performance bottlenecks occur.
{\em \Acl{cb}} thus integrates the systematic performance evaluation on different \ac{hpc} hardware platforms directly into the code development and maintenance process.
Developers and users profit from an easy-accessible and interactive visualization of the performance data. 
To show the portability of the concept, we apply our strategy to two different \ac{hpc} codes,
\fe{}~\cite{MR4318427,million-way} and \wlb{}~\cite{BAUER2021478}, which are fundamentally different in their method, software architecture, and performance characteristics.

The remainder of this paper is structured as follows:
\autoref{sec:background} provides an overview of the two example applications \fe{} and \wlb{} and their underlying methods.
The general concept of \acf{cb} and our realization are described in ~\autoref{sec:cbconcept} and ~\autoref{sec:pipeline}.
\autoref{sec:eval} shows that we could gather meaningful results by implementing our \ac{cb} strategy.
Ultimately,~\autoref{sec:relwork} gives an overview of related work and \autoref{sec:outro} draws a conclusion and gives an outlook for future work.

\section{Background}\label{sec:background}

\subsection{Example Code 1: \fe{}}\label{sec:fe}

\subsubsection{The FE$^2$ Method}

The FE$^2$ method~\cite{FEYEL_FE2_1999,Kouz_Hom_2001,MieheBound,HacklSchroeder,SMIT_Hom_1998} is a nonlinear computational scale bridging approach which efficiently incorporates a microscopic material structure into a macroscopic finite element simulation.
The method is widely used in the field of solid mechanics for small and large deformation processes.
In general, it is assumed that a \ac{rve} is defined, which is a small volume $\mathcal{B}$ that has a microstructure representative for the overall material structure.
The RVE is discretized with finite elements resolving the microstructure.

\begin{figure}
	\centering
	\includegraphics[width=0.7\textwidth]{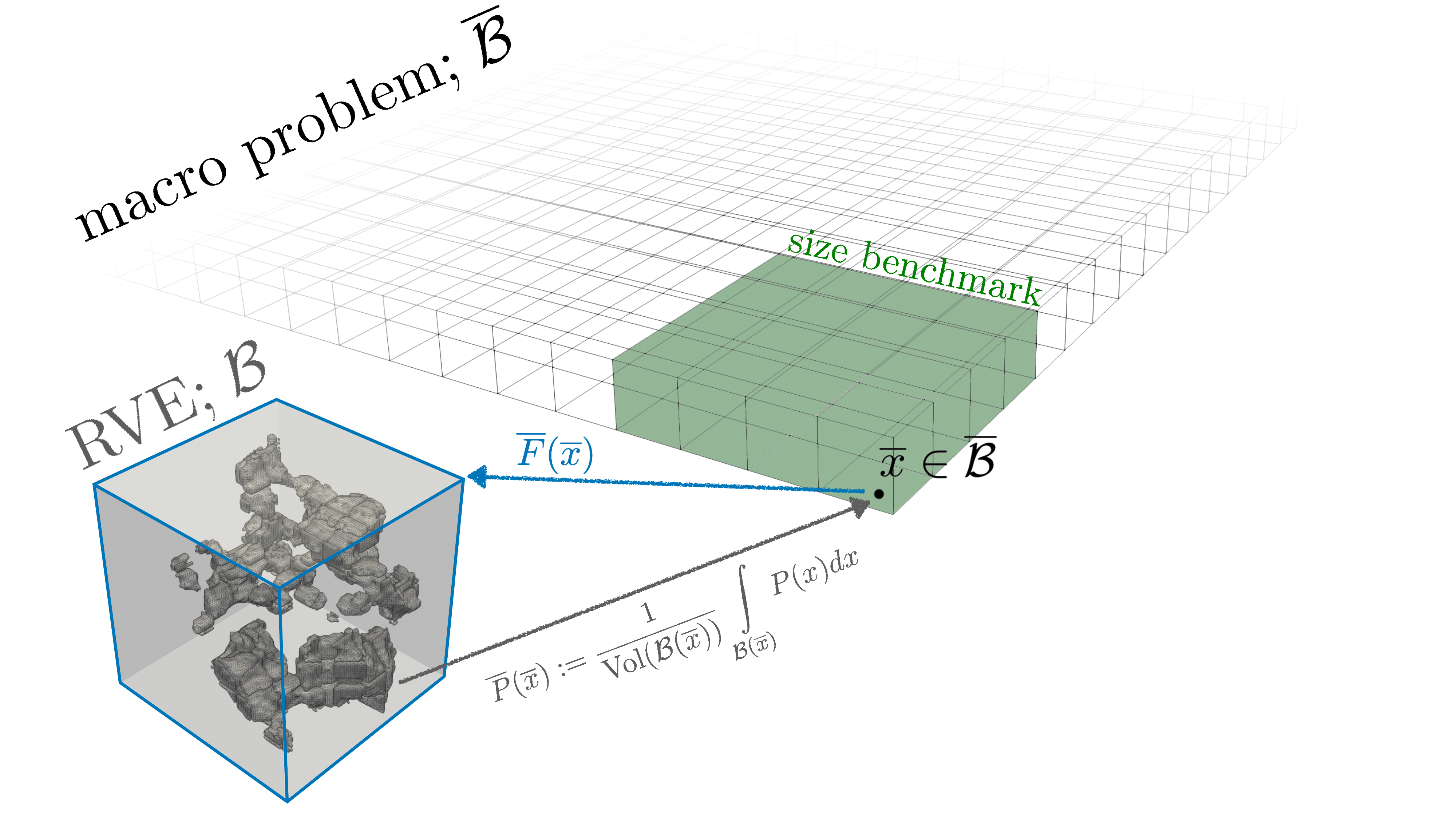}
	\caption{Schematic view of the FE$^2$ approach. The green part of the macroscopic problem in
	shows the size of the part which is handled by one compute node in all benchmarks defined later on, that is, 8 finite elements with 216 attached RVEs.}
	\label{fig:fe2_visu}
\end{figure}

The macroscopic problem $\overline{\mathcal{B}}$ is discretized with finite elements that do not resolve any microstructure, material laws, or material parameters.
Instead, the relation between the current  macroscopic deformation gradient $\overline{F}(\overline{x})$ and the first Piola-Kirchhoff stress tensor $\overline{P}(\overline{x})$ in any integration point $\overline{x}$ of the macroscopic finite elements is computed as follows:
One representative volume element $\mathcal{B}(\overline{x})$ is attached to each integration point and deformed following the macroscopic deformation gradient $\overline{F}(\overline{x})$ in that point.
More precisely, $\overline{F}(\overline{x})$ is used to define a deformation of the attached RVE using, e.g., periodic boundary conditions applied to $\mathcal{B}(\overline{x})$.
Then, a fully nonlinear finite element simulation with appropriate material law and parameters for the different material phases of the microstructure is performed using appropriate boundary conditions.
This is done for all the RVEs in all the integration points of the macroscopic problem.
Finally, the macroscopic stress is defined by a volumetric average over the attached RVE, that is,
$$
	\overline{P}(\overline{x}) := \frac{1}{{\rm Vol}(\mathcal{B}(\overline{x}))}\int\limits_{\mathcal{B}(\overline{x})} P(x) dx,
$$
where $P(x)$ is the microscopic first Piola-Kirchhoff stress tensor.
A schematic view of the FE$^2$ method is shown in Fig.~\ref{fig:fe2_visu}.
There, in an exemplary macroscopic integration point $\overline{x}$, the macroscopic deformation gradient $\overline{F}(\overline{x})$ is used to define the boundary constraints of the RVE $\mathcal{B}$.
After solving the RVE problem, the macroscopic stress $\overline{P}(\overline{x})$ is computed by averaging over the microscopic stresses $P$.
Let us note that in all macroscopic simulations, triquadratic finite elements with 27 integration points are used, and in all these points an RVE is attached.
The computation of the \acp{rve} can happen in parallel.
The green part of the macroscopic problem in Fig.~\ref{fig:fe2_visu} shows the size of the part, which is handled by one compute node in all benchmarks defined later on, that is, 8 finite elements with 216 attached RVEs in total.

\subsubsection{Algorithmic Description}

For a fully nonlinear problem we have to solve a nonlinear problem on the macroscale and many nonlinear problems on the microscale.
Since it is possible to compute the exact tangent or derivative to the residual $\overline{P}$ (see~\cite{HacklSchroeder}), we use Newton's method on the macroscopic scale.
In each macroscopic Newton step, of course, a new state of deformation is reached and to compute the new residual, the stress
$$
	\overline{P}(\overline{x}) := \frac{1}{{\rm Vol}(\mathcal{B}(\overline{x}))}\int\limits_{\mathcal{B}(\overline{x})} P(x) dx
$$
has to be evaluated in all integration points.
Therefore, for each of them a nonlinear RVE problem has to be solved, again with Newton's method.
Since all these RVE problems are completely independent of each other, they can be solved in parallel.
For many critical nonlinear scenarios, it is necessary to perform the macroscopic deformation process in a pseudo-time stepping scheme or load stepping approach, that is, the total deformation is computed in several incremental steps.
Then, each load step is initiated with the solution of the latter one.
As a consequence, the algorithmic structure involves, three nested loops: the outermost loop is the pseudo-time stepping, the second nested loop is the macroscopic Newton's method, the innermost loops are the parallel Newton's methods on the RVEs.

\subsubsection{The FE2TI Implementation}\label{subsec:feimpl}

\fe{}~\cite{MR4318427,million-way} is a scalable implementation of the FE$^2$ method within the PETSc~\cite{petsc-web-page} environment using C/C++.
Originally, it was parallelized using \ac{mpi}~\cite{mpiforumForum}.
Recently, support for pure OpenMP~\cite{openmpHome} and hybrid OpenMP/\ac{mpi} parallelization was added, which allows for a more flexible distribution of the RVEs to the parallel processes and a refined load balance.
The \fe{} software showed its efficiency and scalability for different applications with hyperelastic or elasto-plastic materials, for example, a simulation of the Nakajima test for a dual-phase steel~\cite{MR4318427}.
For the solution of small and medium RVEs, different direct solver packages are supported, as, e.g., UMFPACK~\cite{10.1145/992200.992206}, MUMPS~\cite{mumpssolverMUMPSParallel}, and MKL-PARDISO~\cite{intelOneMKLPARDISO}.
Recently, also simple inexact options have been added where an iterative Krylov subspace solver with a simple preconditioner, as, e.g., \ac{ilu}, is used.
For large RVEs, in former simulations, also (nonlinear) finite element tearing and interconnecting - dual-primal (FETI-DP)~\cite{feti:farhat,feti:nonlinear} domain decomposition methods and algebraic multigrid have been used, which basically introduces a further level of parallelism.

For the solution of the linearized macroscopic problem the same solver options can be used except of FETI-DP.
As an alternative, our state-of-the-art \ac{bddc} implementation~\cite{bddc:klawonn} can be chosen, which is a quite efficient and robust parallel solver for large macroscopic problem sizes.

For the different benchmark problems used and defined in this paper, we always consider simple deformations of dual-phase steel blocks.
All RVEs have a simple spherical inclusion of martensite in a ferritic matrix phase and a $J_2$ elasto-plasticity material model is used.
The material parameters for martensite and ferrite are taken from~\cite{Brands2016} and the implementation of the material model in FEAP is taken from~\cite{klinkel2000theorie}.
We use this setup in two different variations.
The first one uses a small macroscopic cube with $2\times2\times2$ triquadratic finite elements and therefore 216 \acp{rve}.
The total deformation of 0.025\% is applied in 2 load steps.
The size of the RVE can range from 1296 to 6000 quadratic tetrahedral finite elements, that is, 6591 to 27783 degrees of freedom.
We will refer this benchmark as \textit{fe2ti216}.
The second one, uses a macroscopic cube size of $8\times8\times1$ triquadric finite elements and 1728 \acp{rve}, we will refer to that benchmark as \textit{fe2ti1728}.

As sparse direct solvers, we use in this paper MKL-PARDISO and UMFPACK, as well as an inexact option, GMRES with ILU preconditioner.
For the sake of simplicity, we will use PARDISO to refer to MKL-PARDISO and ILU to refer to GMRES with the ILU preconditioner.

\subsection{Example Code 2: \wlb{}}\label{sec:walberla}

\subsubsection{\acl{lbm}}\label{subsec:lbm}
\acf{lbm} is a modern computational fluid dynamics approach~\cite{kruger2017LatticeBoltzmannMethod}.
It discretizes the Boltzmann equation on a Cartesian lattice with spacing $\Delta x \in \mathbb{R^{+}}$ using D$d$Q$q$ velocity sets.
\Acp{pdf} $f_{i}(\boldsymbol{x}, t) \in \mathbb{R}$ represent density probability of the fluid particles.
The discrete LBM equation involves collision 
\begin{equation}\label{eq:nm-lbm-collision}
	f_{i}^{\star}(\boldsymbol{x}, t) = f_{i}(\boldsymbol{x}, t) + \Omega_{i}(\boldsymbol{x}, t) + F_{i}(\boldsymbol{x}, t)
\end{equation}
and streaming 
\begin{equation}\label{eq:nm-lbm-streaming}
	f_{i}(\boldsymbol{x} + \boldsymbol{c}_{i}\Delta t, t+\Delta t) = f_{i}^{\star}(\boldsymbol{x}, t).
\end{equation}
steps, where $\Omega_{i}(\boldsymbol{x}, t)$ describes the collision operator, $F_{i}(\boldsymbol{x}, t)$ external forces and $f_{i}^{\star}(\boldsymbol{x}, t)$ the equilibrium.

For the collision, a Single Relaxation Time (SRT) operator 
\begin{equation}\label{eq:nm-lbm-collision-operator}
	\Omega_{i}(\boldsymbol{x}, t) = \frac{f_{i}(\boldsymbol{x}, t) - f_{i}^{\text{eq}}(\boldsymbol{x}, t)}{\tau} \Delta t
\end{equation}
is employed with relaxation time $\tau > \Delta t / 2$.
The SRT describes the basic collision operator~\cite{kruger2017LatticeBoltzmannMethod}.
Other collision operator incldue the two-relaxation time (TRT), multiple-relaxation time (MRT), central moments, cumulants, and entropic collision operators.
The collision operator used is therefore a parameter that can be adjusted to fit the application~\cite{PhysRevE.100.033305}. 

The equilibrium 
\begin{equation}\label{eq:nm-lbm-equilibrium}
	f_{i}^{\text{eq}}(\boldsymbol{x}, t) = w_{i}\rho\left(1 + \frac{\boldsymbol{u}\cdot\boldsymbol{c}_{i}}{c_{s}^{2}} + \frac{(\boldsymbol{u} \cdot \boldsymbol{c}_{i})^{2}}{2 c_{s}^{4}} - \frac{\boldsymbol{u} \cdot \boldsymbol{u}}{2 c_{s}^{2}} \right).
\end{equation}
is derived from the Maxwell--Boltzmann equation. Typical stencils are D$2$Q$9$, D$3$Q$19$ or D$3$Q$27$ lattice models are used with weights $w_{i}$ and speed of sound $c_{s}^{2}$.

The density 
\begin{equation}\label{eq:nm-lbm-density}
	\rho(\boldsymbol{x}, t) = \sum_{i} f_{i}(\boldsymbol{x}, t)
\end{equation}
and velocity 
\begin{equation}\label{eq:nm-lbm-velocity}
	\boldsymbol{u}(\boldsymbol{x}, t) = \frac{\boldsymbol{F}(\boldsymbol{x}, t)\Delta t}{2 \rho(\boldsymbol{x}, t)} + \frac{1}{\rho(\boldsymbol{x}, t)}\sum_{i} \boldsymbol{c}_{i} f_{i}(\boldsymbol{x}, t),
\end{equation}
can be computed from moments of \acp{pdf}.
The kinematic viscosity 
\begin{equation}\label{eq:nm-lbm-viscosity}
	\nu = c_{s}^{2} \left(\tau - \frac{\Delta t}{2}\right)
\end{equation}
is related to relaxation time $\tau$. External forces, like gravity, are modeled with 
\begin{equation}\label{eq:nm-lbm-force-guo}
	F_{i}(\boldsymbol{x}, t) =
	\left(1 - \frac{\Delta t}{2 \tau} \right)
	w_{i}\left(\frac{\boldsymbol{c}_{i} - \boldsymbol{u}}{c_{s}^{2}} + \frac{(\boldsymbol{c}_{i} \cdot \boldsymbol{u}) \boldsymbol{c}_{i}}{c_{s}^4}\right) \cdot \boldsymbol{F}(\boldsymbol{x}, t).
\end{equation}

The boundary conditions include bounce-back at solid obstacles and specular reflection for free-slip.
Common LBM units with $\Delta x=1$ and $\Delta t=1$ are assumed.
Reference density $\rho_{0}=1$ and pressure $p_{0} = c_{s}^{2} \rho_{0} = 1/3$ are set in all simulations.
A common metric for evaluating the performance of an implementation of \ac{lbm} is the number of \ac{mlups}.
This refers to the rate of lattice updates, i.e. one streaming and one collision operation, performed per second.

\subsubsection{Free Surface Lattice Boltzmann Method}\label{subsec:fslbm}

The \ac{fslbm} in this work uses the implementation of Schwarzmeier et al.~\cite{Schwarzmeier2022, Schwarzmeier2022Refilling, Schwarzmeier2023}.
It simulates a dynamic interface between two immiscible fluids, where the heavier fluid governs flow dynamics.
This reduces the problem to a single-fluid flow with a free boundary, valid for significantly different fluid densities and viscosities.
The heavier fluid can be interpreted as liquid, and the lighter one as gas.

The interface uses the volume-of-fluid approach~\cite{hirt1981VolumeFluidVOF}, assigning a fill level $\varphi (\boldsymbol{x}, t)$ to each lattice cell, indicating the phase of the cell.
Cells can be liquid ($\varphi(\boldsymbol{x}, t)=1$), gas ($\varphi(\boldsymbol{x}, t)=0$), or interface ($\varphi(\boldsymbol{x}, t) \in \left(0, 1\right)$).
Liquid, gas, and interface cells are treated differently in the \ac{fslbm} simulation.
The liquid mass of a cell is determined by
\begin{equation}\label{eq:nm-fslbm-mass}
	m\left(\boldsymbol{x}, t\right) = \varphi\left(\boldsymbol{x}, t\right) \rho\left(\boldsymbol{x}, t\right) \Delta x^{3}
\end{equation}
with $\rho\left(\boldsymbol{x}, t\right)$ fluid density, and $\Delta x^{3}$ volume.
The mass flux between the interface and other cells is computed during streaming with

\begin{equation}\label{eq:nm-fslbm-mass-flux}
	\resizebox{\textwidth}{!}{%
		$\frac{\Delta m_{i}\left(\boldsymbol{x}, t\right)}{\Delta x^{3}} =
			\begin{cases}
				0                                                                                                                                 & \boldsymbol{x} + \boldsymbol{c}_{i}\Delta t \in \text{gas}        \\

				f_{\overline{i}}^{\star}\left(\boldsymbol{x} + \boldsymbol{c}_{i}\Delta t, t\right) - f_{i}^{\star}\left(\boldsymbol{x}, t\right) & \boldsymbol{x} + \boldsymbol{c}_{i}\Delta t \in \text{liquid}     \\

				\frac{1}{2}\Bigl(\varphi\left(\boldsymbol{x}, t\right) + \varphi\left(\boldsymbol{x} + \boldsymbol{c}_{i}\Delta t, t\right) \Bigr)
				\Bigl(f_{\overline{i}}^{\star}\left(\boldsymbol{x} + \boldsymbol{c}_{i}\Delta t, t\right) -
				f_{i}^{\star}\left(\boldsymbol{x}, t\right)\Bigr)                                                                                 & \boldsymbol{x} + \boldsymbol{c}_{i}\Delta t \in \text{interface},
			\end{cases}$
	}
\end{equation}
with $\bar{i}$ defining the reverse lattice direction resulting in $\boldsymbol{c}_{\overline{i}} = -\boldsymbol{c}_{i}$.

Interface cell conversions via
\begin{equation}\label{eq:nm-fslbm-excess-mass}
	\frac{m_{\text{ex}}\left(\boldsymbol{x}, t\right)}{\rho\left(\boldsymbol{x}, t\right) \Delta x^{3}} =
	\begin{cases}
		\varphi^{\text{conv}}\left(\boldsymbol{x}, t\right) - 1 & \text{if } \boldsymbol{x} \text{ is converted to liquid} \\
		\varphi^{\text{conv}}\left(\boldsymbol{x}, t\right)     & \text{if } \boldsymbol{x} \text{ is converted to gas}.
	\end{cases}
\end{equation}
are controlled by a threshold ($\varepsilon_{\varphi}=10^{-2}$) to prevent oscillatory conversions~\cite{pohl2008HighPerformanceSimulation}.
The excess mass is evenly distributed among neighboring interface cells during conversions to conserve total mass.
Unnecessary interface cells without gas or liquid neighbors are adjusted in mass flux.

When converting cells, there is no modification of the PDFs.
In gas-to-interface conversions, PDFs are initialized based on the equilibrium~\eqref{eq:nm-lbm-equilibrium}.
The collision~\eqref{eq:nm-lbm-collision} and streaming~\eqref{eq:nm-lbm-streaming} occur in interface and liquid cells.

The macroscopic boundary condition~\cite{scardovelli1999DirectNumericalSimulation,bogner2017DirectNumericalSimulation} at the free surface modeled by
\begin{equation}\label{eq:nm-fslbm-boundary-condition-macroscopic}
	\begin{aligned}
		p\left(\boldsymbol{x}, t\right) - p^{\text{G}}\left(\boldsymbol{x}, t\right) + p^{\text{L}}\left(\boldsymbol{x}, t\right) & =  2\mu\partial_{n}u_{n}\left(\boldsymbol{x}, t\right)                                                       \\
		0                                                                                                                         & = \partial_{t_{1}}u_{n}\left(\boldsymbol{x}, t\right) + \partial_{n}u_{t_{1}}\left(\boldsymbol{x}, t\right)  \\
		0                                                                                                                         & = \partial_{t_{2}}u_{n}\left(\boldsymbol{x}, t\right) + \partial_{n}u_{t_{2}}\left(\boldsymbol{x}, t\right).
	\end{aligned}
\end{equation}
with gas pressure $p^{\text{G}}\left(\boldsymbol{x}, t\right)$, Laplace pressure $p^{\text{L}}\left(\boldsymbol{x}, t\right)$, tangent vectors $\boldsymbol{t}_{1}(\boldsymbol{x}, t) \in \mathbb{R}^{d}$ and $\boldsymbol{t}_{2}(\boldsymbol{x}, t) \in \mathbb{R}^{d}$.
At the free surface interface, the LBM anti-bounce-back pressure boundary condition~\cite{korner2005LatticeBoltzmannModel}
\begin{equation}\label{eq:nm-fslbm-boundary-condition}
	f_{i}^{\star}\left(\boldsymbol{x} - \boldsymbol{c}_{i}\Delta t, t\right)
	= f_{i}^\text{eq}\left(\rho^{\text{G}}\left(\boldsymbol{x}, t\right), \boldsymbol{u}\left(\boldsymbol{x}, t\right)\right)
	+ f_{\overline{i}}^\text{eq}\left(\rho^{\text{G}}\left(\boldsymbol{x}, t\right), \boldsymbol{u}\left(\boldsymbol{x}, t\right)\right)
	- f_{\overline{i}}^{\star}\left(\boldsymbol{x}, t\right) 
\end{equation}
is used.
Here $\rho^{\text{G}}\left(\boldsymbol{x}, t\right)$ defines the gas density and $\boldsymbol{u}\left(\boldsymbol{x}, t\right)$ the velocity of the free surface interface.
Due to their unavailability, all PDFs streaming from gas cells to interface cells must adhere to the free-surface boundary condition~\eqref{eq:nm-fslbm-boundary-condition}.

The gas pressure is expressed as
\begin{equation}\label{eq:nm-fslbm-pressure-gas}
	p^{\text{G}}\left(\boldsymbol{x}, t\right) = p^{\text{V}}\left(t\right) - p^{\text{L}}\left(\boldsymbol{x}, t\right),
\end{equation}
where $p^{\text{V}}(t)$ represents volume pressure, and $p^{\text{L}}(\boldsymbol{x}, t)$ is Laplace pressure.
The Laplace pressure is determined by surface tension $\sigma \in \mathbb{R^{+}}$ and interface curvature $\kappa(\boldsymbol{x}, t) \in \mathbb{R}$
\begin{equation}\label{eq:nm-fslbm-pressure-laplace}
	p^{\text{L}}\left(\boldsymbol{x}, t\right) = 2 \sigma \kappa\left(\boldsymbol{x}, t\right).
\end{equation}
In this study, the interface curvature is calculated using finite difference methods \cite{bogner2016CurvatureEstimationVolumeoffluid}
\begin{equation}
	\kappa(\boldsymbol{x}, t) = -\nabla \cdot \boldsymbol{\hat{n}}(\boldsymbol{x}, t),
\end{equation}
where the normalized interface normal is obtained through a weighted central finite difference methods\cite{parker1992TwoThreeDimensional}
\begin{equation} \label{eq:nm-fslbm-normal}
	\boldsymbol{n}(\boldsymbol{x}, t) = \nabla \varphi(\boldsymbol{x}, t).
\end{equation}
The computation of $\boldsymbol{n}(\boldsymbol{x}, t) \in \mathbb{R}^{d}$ is adjusted near solid obstacle cells~\cite{donath2011WettingModelsParallel}.
The curvature $\kappa(\boldsymbol{x}, t)$ is effectively computed from the second-order derivative of the fill level $\varphi(\boldsymbol{x}, t)$.
To mitigate errors introduced by the non-smooth indicator function $\varphi(\boldsymbol{x}, t)$, a smoothing process using the \textit{K\textsubscript{8}}-Kernel~\cite{williams1999AccuracyConvergenceContinuum} with a support radius of 2.0 is applied~\cite{bogner2016CurvatureEstimationVolumeoffluid}.

\subsubsection{The \wlb{} Implementation}\label{subsec:wlb}

\wlb~\cite{BAUER2021478, chair_for_system_simulation_2023_10054460} is a multiphysics HPC C++ framework for simulations on massively parallel systems.
\wlb~initially targeted LBM simulations but evolved into a general-purpose multiphysics framework, including rigid particle dynamics, phase-field simulations, and the coupled algorithms like the \ac{fslbm}.
It utilizes fully distributed data structures in uniform blocks to ensure scalability, supporting pure \ac{mpi} and hybrid MPI/OpenMP parallelization.
Designed with modularity in mind, \wlb{}, together with its code generation extension \emph{lbmpy}~\cite{BAUERlbmpy, HENNIGlbmpy}, enhances productivity, reusability, and maintainability.
Using metaprogramming techniques, the code generation python package \emph{lbmpy} facilitates the efficient implementation and automated generation of different \ac{lbm} variants, resulting in highly optimized compute kernels for different hardware.
With these techniques, switching between different implemented stencils, streaming patterns, and collision operators is easily possible.
\emph{lbmpy} supports a range of different collision operators, e.g.~the SRT, two-relaxation time (TRT), multiple-relaxation time (MRT), central moments, cumulants, and entropic collision operators.

\begin{figure}
	\centering
	\setlength{\figureheight}{0.40\textwidth}
	\setlength{\figurewidth}{0.40\textwidth}
	\setlength\mdist{0.02\textwidth}
	\begin{tikzpicture}
	
	\begin{axis}%
		[width=\figurewidth,
		height=\figureheight,
		xmin=0,
		xmax=1,
		ymin=0,
		ymax=1,
		ticks=none,
		axis lines=none,
		clip=false,
		scale only axis
		]
		\addplot[thick, name path=f, domain=0:1,samples=50,smooth,blue] {0.5+0.125*cos(deg(pi*x*2))};
		
		\path[name path=axis] (axis cs:0,0) -- (axis cs:1,0);
		
		\addplot [thick, color=blue, fill=blue!30] fill between[of=f and axis, soft clip={domain=0:1}];
	\end{axis}
	
	\draw[very thick, loosely dotted, black!60] (0,0)--(0,\figureheight);
	\draw[very thick, loosely dotted, black!60] (\figurewidth,0)--(\figurewidth,\figureheight);
	
	\draw[very thick,, black!50] (0,\figureheight)--(\figurewidth,\figureheight);
	\draw[very thick,, black!50] (0,0)--(\figurewidth,0);
	
	\draw[<->, >=Latex] (\figurewidth+\mdist,0)--(\figurewidth+\mdist,\figureheight) node [pos=0.5,right] {$l$};
	\draw[-] (\figurewidth,0)--(\figurewidth+1.5\mdist,0);
	\draw[-] (\figurewidth,\figureheight)--(\figurewidth+1.5\mdist,\figureheight);
	
	\draw[<->, >=Latex] (0,-\mdist)--(\figurewidth,-\mdist) node [pos=0.5,below] {$l$};
	\draw[-] (0,0)--(0,-1.5\mdist) node [below] {$x=0$};
	\draw[-] (\figurewidth,0)--(\figurewidth,-1.5\mdist);
	
	\draw[dashed, gray] (0,0.5\figureheight)--(\figurewidth,0.5\figureheight);
	\draw[<->, >=Latex] (-\mdist,0)--(-\mdist,0.5\figureheight) node [pos=0.5,left] {$h$};
	\draw[-] (-1.5\mdist,0)--(0,0);
	\draw[-] (-1.5\mdist,0.5\figureheight)--(0,0.5\figureheight);
	
	\draw[<->, >=Latex] (-\mdist,0.5\figureheight)--(-\mdist,0.625\figureheight) node [pos=0.5,left] {$a_{0}$};
	\draw[-] (-1.5\mdist,0.625\figureheight)--(0,0.625\figureheight);
	
	\node[
	rectangle,
	anchor=south,
	blue] at (0.5\figurewidth,0.57\figureheight) {$y(x) = h + a_{0} \cos\left(kx \right)$};
	
	
	\draw[thick, ->, >=Latex, red] (.5\figurewidth,.5\figureheight-3\mdist)--(.5\figurewidth,.5\figureheight-6\mdist) node [pos=0.5,right] {$g$};
	
	\draw[->, >=Latex] (2\mdist,2\mdist)--(4\mdist,2\mdist) node [below] {$x$};
	\draw[->, >=Latex] (2\mdist,2\mdist)--(2\mdist,4\mdist) node [left] {$y$};

\end{tikzpicture}
	\caption{Illustrates the initialization of the gravity wave, using the fluid depth $h$, initialized amplitude of the wave $a_0$, wavenumber $k = 2 \pi / l$, wavelength $l$ and gravitational acceleration $g$. In $x$ and $z$-direction periodic and no slip boundary conditions in $y$-direction were used (based on \cite{Schwarzmeier2022})}\label{fig:gravityWave}
\end{figure}
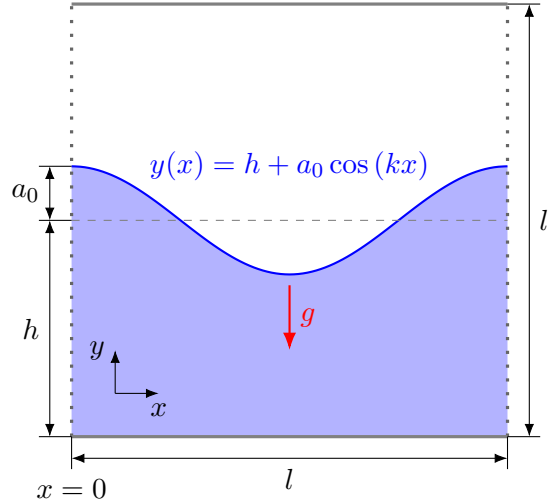

We use a plain \ac{lbm} benchmark that tests different collision operators on a uniform grid.
The benchmark is executed on both CPUs and GPUs and is referred to as \textit{UniformGridCPU} for the CPU variant or \textit{UniformGridGPU} for the GPU variant.
Next to this plain \ac{lbm} benchmark case, there is also a benchmark case for the free-surface lattice Boltzmann method (c.f. \autoref{subsec:fslbm}).
The benchmark employed in this study emulates a gravity wave, as depicted in~\autoref{fig:gravityWave}.
Therefore, we will refer to this benchmark case as \textit{GravityWaveFSLBM}.
Since our current \ac{fslbm} implementation only runs on CPUs, there is currently no GPU case for that.
Prior investigations have leveraged this and similar test cases to verify the physical accuracy of the FSLBM implementation~\cite{Schwarzmeier2022, Schwarzmeier2022Refilling, Schwarzmeier2023}.
Furthermore, it serves as a suitable benchmark for performance analysis, aiming to establish equal computational loads across all computational blocks.
This is achieved through an initialization procedure wherein each block receives a gravity wave.
Here, periodic boundary conditions in $x$ and $z$-direction and no-slip boundary conditions in $y$-direction are used.
Mesh refinement and load balancing algorithms are omitted to optimize performance evaluation of the FSLBM in favor of strict uniformity in load distribution, given that block decomposition is solely conducted in the $x$ and $z$ directions.
As a result of this initialization, each computational block encompasses all three distinct cell states — fluid, gas, and interface.

\begin{table}
	\caption{Comparison between the two example codes.}\label{tab:comp}
	\begin{center}
		\begin{tabular}[c]{lcc}
			\toprule
			                                            & \fe{}                 & \wlb{}                          \\
			\midrule
			\multirow{2}{*}{Field}                      & material science,     & \multirow{2}{*}{fluid dynamics} \\
			                                            & homogenization        &                                 \\
			Language                                    & C/C++                 & C/C++                           \\
			Algorithm                                   & FE$^2$                & \ac{lbm}                        \\
			Solver                                      & implicit              & explicit                        \\
			Software architecture                       & PETSc-based           & Framework                       \\
			\multirow{2}{*}{Performance critical parts} & RVE solver            & handwritten or                  \\
			                                            & (direct or iterative) & generated kernels               \\
			\multirow{2}{*}{Parallelization}            & MPI/Hybrid            & MPI/Hybrid                      \\
			                                            & (with OpenMP)         & (with OpenMP)                   \\
			Accelerators                                & -                     & GPUs                            \\
			Build tool                                  & Make                  & CMake                           \\
			\bottomrule
		\end{tabular}
	\end{center}
\end{table}

\section{From Continuous Integration to Continuous Benchmarking}\label{sec:cbconcept}


\Acl{ci} is a common practice often among software engineers.
It involves teams of developers regularly contributing their changes to a shared codebase.
Each contribution is integrated into the source code and the software is then automatically built, checked, and tested.
This ensures that errors are detected more quickly, and prevents each developer from having their own version of the software~\cite{fowler2006continuous}.
In practice, developers submit their code changes to a shared repository of a \acl{vcs}.
This keeps track of all the changes and versions of the source code, and helps coordinate the integration of new contributions.
Git~\cite{gitscm} is a widely used \acl{vcs}, and platforms such as \gl{}~\cite{gitlab} or \textit{GitHub}~\cite{github} offer hosted Git repositories as a web service.
They also provide a graphical user interface in a web browser, bug trackers, and other features that make it easier for teams and communities to work together.
All of these platforms support \acl{ci} workflows so that each new version of the software can be automatically built, tested, and deployed.

In scientific or \ac{hpc} software, these software engineering principles have been used less frequently. 
There are various reasons for this, one of which could be that scientific software developers have an application background rather than a software engineering background.
This changed in the last decades as communities and initiatives like
\textit{\ac{bssw}}\cite{bsswio}, \textit{de-RSE}\cite{derse} or \textit{SURESOFT}~\cite{blech_2023_7737560}
advocate the use of the software engineering techniques for scientific software~\cite{Anzt2021} under the term \acl{rse}.
Another initiative that has similar goals is \ac{xSDK}~\cite{bartlett2017xsdk}, which defines common interfaces, conventions and best practices
for extreme-scale scientific software.
This should provide the infrastructure that helps \ac{hpc} software developers to write more sustainable, portable, maintainable, and interoperable codes.
Well-known software packages that are part of the consortium are PETSc\cite{petsc-web-page}, Trilinos\cite{trilinos-website}, Hypre\cite{hypre}, and the GINKGO library\cite{ginkgo-toms-2022}.
Thus, nowadays the use of \acl{ci} tools that are offered by the \acl{vcs} platforms is widely used for scientific software packages.

As mentioned in the introduction, computational performance is an important aspect of scientific software used for large-scale simulations.
In some cases, it is also the aspect that is the subject of research.
Continuous testing of performance in the target environment is also important.
Thus, the concept of \acl{ci} has been extended to \acl{cb}.

The concept of \ac{cb} is visualized in \autoref{fig:concept}.
Similar to other \acl{ci} applications, it means that with every change tracked by the \acl{vcs}, the code gets automatically benchmarked.
To get meaningful results that yield insights about the production use of the software, these automatic benchmarks should run on \iac{hpc} system.
Various metrics, such as timings and hardware performance counters, are collected for analysis.
These metrics are subsequently stored for later review and comparison.
Finally, they are visualized to provide feedback to the developers.

With our \ac{cb} pipeline, which we present in the \autoref{sec:pipeline}, we aim to reduce the overhead for developers when running benchmarks on a \ac{hpc} cluster.
Therefore, we automate the benchmarking of specific code revisions on different hardware architectures or parameter configurations.
This would be tedious and time-consuming if done by manually.
Additionally, the collected results are stored in a structured manner and are visualised interactively.
In the short term, this provides developers with prompt feedback on how a code change impacts performance across various architectures or configurations.
In the long term, this allows for tracking of the evolution of performance and tracing of the impact of code changes on performance characteristics.

\begin{figure}
	\centering
	\begin{minipage}[t]{0.49\textwidth}
		\resizebox{\columnwidth}{!}{
%
%

\begin{tikzpicture}[
    -latex,
    auto,
    on grid,
    thick,
    node distance=80pt,
    item/.style ={
        rectangle,
        top color=white,
        draw,
        text=black,
        text centered,
        thick,
        align=center,
        minimum width=80pt,
        minimum height=25pt,
        rounded corners=4pt
    }
]

\node[item] (A) {Source Code};
\node[item] (B) [below right=of A] {HPC Resources};
\node[item] (C) [below left=of B] {Data Storage};
\node[item] (D) [below left=of A] {Developer/User};

\path[draw, thick, -latex, bend left=30] (A) edge node[midway, right, xshift=5pt] {Submit Benchmarks} (B);
\path[draw, thick, -latex, bend left=30] (B) edge node[midway] {Store Results} (C);
\path[draw, thick, -latex, bend left=30] (C) edge node[midway, left, xshift=-5pt] {Visualization} (D);
\path[draw, thick, -latex, bend left=30] (D) edge node[midway] {Commit Code Changes} (A);

\end{tikzpicture}
		}
		\caption{Concept of the \ac{cb} pipeline. }\label{fig:concept}
	\end{minipage}
	\begin{minipage}[t]{0.49\textwidth}
		\resizebox{\columnwidth}{!}{
			\input{figures/cbpipeline.tex}
		}
		\caption{Implementation of the \ac{cb} pipeline.}\label{fig:impl}
	\end{minipage}
\end{figure}

\section{The Continuous Benchmarking Pipeline}\label{sec:pipeline}

This section describes the actual realization of the \ac{cb} pipeline.
First, the different components are described individually, and the last part of the section presents the implementations for our two example applications.
\autoref{fig:impl} shows how the components play together to form the complete \ac{cb} pipeline.

\subsection{Hardware/Software environment}\label{subsec:env}

Our pipeline aims to run the benchmarks on various hardware architectures and track the single node performance.
Even so, the example codes are meant to run on many nodes; a good single-node performance is the basis for an excellent multi-node performance.
Therefore, we use the Testcluster~\cite{fauTestclusterNHRFAU}, a special \ac{hpc} resource of the \nhr{}.
It is meant for testing and benchmarking and consists of different compute nodes where each node uses different CPUs or GPUs (c.f.~\autoref{tab:server-configurations}).
As all the nodes in that cluster are from a different architecture, it only allows running single-node jobs.
In contrast to other compute centers, no generic \acl{ci} driver like jacamar~\cite{gitlabExascaleComputing} is used, but a custom solution \gl{} runner.
With this custom \gl{} runner that is accessible from the \gl{} instances offered by the \nhr{} it is possible to use the \gl{} \acl{ci} features and execute jobs on the Testcluster~\cite{fauOverviewNHRFAU}.
When a commit is pushed to the code repository, the pipeline is triggered, and the software is built and executed on selected nodes on the Testcluster.

\begin{table}
	\centering
	\caption{Excerpt of the available compute nodes in the Testcluster at \nhr{}~\cite{fauTestclusterNHRFAU}}
	\resizebox{\textwidth}{!}{%
		\begin{tabular}{llll}
			\toprule
			\textbf{Hostname}                & \textbf{CPU}                                                  & \textbf{\#Cores}              & \textbf{Accelerators}         \\
			\midrule
			\texttt{casclakesp2}             & Dual Intel Xeon "Cascade Lake" Gold 6248 CPU                  & 2x 20 cores                   &                               \\
			\texttt{euryale}                 & Dual Intel Xeon "Broadwell" CPU E5-2620 v4                    & 2x 8 cores                    & AMD RX 6900 XT                \\
			\multirow{2}{*}{\texttt{genoa2}} & \multirow{2}{*}{Dual AMD EPYC 9354 "Genoa" CPU}               & \multirow{2}{*}{2x 32 cores } & Nvidia A40                    \\
			                                 &                                                               &                               & Nvidia L40s                   \\
			\texttt{hasep1}                  & Dual Intel Xeon "Haswell" E5-2695 v3 CPU                      & 2x 14 cores                   &                               \\
			\texttt{icx36}                   & Dual Intel Xeon "Ice Lake" Platinum 8360Y CPU                 & 2x 36 cores                   &                               \\
			\texttt{ivyep1}                  & Dual Intel Xeon "Ivy Bridge" E5-2690 v2 CPU                   & 2x 10 cores                   &                               \\
			\multirow{4}{*}{\texttt{medusa}} & \multirow{4}{*}{Dual Intel Xeon "Cascade Lake" Gold 6246 CPU} & \multirow{4}{*}{2x 12 cores } & Nvidia Geforce RTX 2070 SUPER \\
			                                 &                                                               &                               & Nvidia Geforce RTX 2080 SUPER \\
			                                 &                                                               &                               & Nvidia Quadro RTX 5000        \\
			                                 &                                                               &                               & Nvidia Quadro RTX 6000        \\
			\texttt{naples1}                 & Dual AMD EPYC 7451 "Naples" CPU                               & 2x 24 cores                   &                               \\
			\texttt{optane1}                 & Dual Intel Xeon "Ice Lake" Platinum 8362 CPU                  & 2x 32 cores                   &                               \\
			\texttt{rome1}                   & Single AMD EPYC 7452 "Rome" CPU                               & 1x 32 cores                   &                               \\
			\texttt{skylakesp2}              & Intel Xeon "Skylake" Gold 6148 CPU                            & 2x 20 cores			 &                               \\
			\bottomrule
		\end{tabular}%
	}
	\label{tab:server-configurations}
\end{table}

\subsection{Job Submission/Execution}

When using the \gl{} \acl{ci} features, the jobs that should be executed are specified in \yml{}~\cite{ben2009yaml} files.
These contain the statements to execute and configuration parameters for each job.
The basic concept of the pipeline is to use the \gl{} runner to assemble or parameterize job scripts, which are then submitted via the installed workload manager.
\autoref{lst:submitjob} shows an exemplary \gl{} \acl{ci} job specification to do so.
Lines 11 and 12 in \autoref{lst:submitjob} shows how the job script is assembled from a basis part, defined in \verb|basic_config.sh|,
containing cluster-specific environment variables and parameters for the batch scheduler
Furthermore, a benchmark-specific part, defined in the variable \verb|SCRIPT|, contains the actual benchmark execution instructions.
A script is available for each benchmark to be executed, and this job is instantiated on each available host for each benchmark.

Instead of directly specifying the executed commands within the \yml{} job specification, which would also be possible, this yields various benefits.
On the one hand, it simplifies the development process of the pipeline, as the job script can be written like a standard job script for the batch scheduler and can be tested without triggering the \gl{} runner.
On the other hand, the job scripts can also be reused in other scenarios.
For example, the same job script can be used for a large-scale run, with minor changes like loading the correct modules.
Another benefit is that it allows the reuse of existing job scripts (with minor adaptions) as a template for the pipeline without translating it into \yml{} files.

During execution, the performance metrics are gathered with the \lwd{} toolsuite~\cite{likwid} for performance metrics executed on the CPUs.
For benchmarks on Nvidia GPUs, the vendor tool \ncu{} is used.

For the benchmark cases in the \ac{cb} setup, we have two special requirements: they should not take too long to complete, but they also should be representative of what happens during a production run.
Additionally, there is the requirement of the used cluster that allows only single-node runs,
the benchmark cases should run on a single node and represent what a single node would do in a large run.
Further, they should produce valuable insights to steer the development process.

\begin{figure}
	\begin{small}
		\begin{center}
			\begin{lstlisting}[language=yaml,label={lst:submitjob},caption={GitLab CI job specification for job that submits a job script via the slurm interface.}]
.submit_job:
  tags:
    - testcluster
  variables:
    NO_SLURM_SUBMIT: 1
    SLURM_TIMELIMIT: 120
    HOST: "TOBEREPLACED"
    SCRIPT: "TOBEREPLACED"
  script:
    - JOB_SCRIPT_FILE="job_script_${HOST}.sh"
    - ./base_config.sh > ${JOB_SCRIPT_FILE}
    - cat "${SCRIPT}" >> ${JOB_SCRIPT_FILE}
    - job_id=$(sbatch --parsable --wait \
      --nodelist="${HOST}" \
      --job-name "${CI_JOB_NAME}" \
      ${JOB_SCRIPT_FILE})
    - cat ${CI_JOB_NAME}.o${job_id}.log
\end{lstlisting}
		\end{center}
	\end{small}
\end{figure}

\subsection{Data Storage}

At the end of a run, the program's output and the output from \lwd{} or \ncu{} are collected and parsed.
The metrics of interest are extracted from the raw output and uploaded to a database.
In the current setup, an instance of the \idb{}~\cite{influxdataInfluxDBRealtime} is used to store the data.
\idb{} is a \ac{tsdb}, i.e., a type of database designed for storing time series and timestamped data and is optimized to track changes over time~\cite{bader2017survey}.
We use the time when the corresponding pipeline is triggered as a timestamp for the data points.
\idb{} distinguishes between fields and tags. Fields are the actual data, the collected runtime metrics, like the \ac{tts}, the number of \ac{flop}, or data traffic measurements.
Tags are the metadata, in our case, we use the program parameter, like the domain size, the direct solver used, or the compute node used.

For reproducibility, the state of the used compute node is tracked with \ms{}\cite{maschinestate}, a tool that collects information about the software and the hardware in a text file.
All the raw output files, like the output files from \lwd{} or \ms{}, the log files from the batch scheduler are then uploaded to a \kadi{}\cite{Brandt2021} instance.
Kadi4Mat is the Karlsruhe Data Infrastructure for Materials Science, an open-source software for managing research data.
It aims to combine the possibilities of structured storage, management and exchange of scientific data based on the \ac{fair}\cite{draxl2018nomad} principles, with documented and reproducible workflows for data analysis, visualization and other tasks while incorporating new concepts as well as existing solutions.
Thus, Kadi4Mat can be observed as logically divided into two components - a repository and an electronic lab notebook for creation and management of workflows.
In this work, Kadi4Mat is utilized for structured data storage, linking of resources and exchange of data with specific research groups/sub-groups.
Kadi4Mat offers various resources for data storage and the simplest ones of them are - the records.
A record can link arbitrary data with descriptive metadata and serve as basic components that can be used in workflows and future data publications.
It can be used for all kinds of data, including simulation or experimental data and it can be linked to other records of related datasets with an optional name describing the linked relationship between records.
A logical grouping of records, for better organization, is referred to as a collection.
A collection can have multiple child collections within itself, providing an option to segregate records on the collection-level.
Records and collections are created and shared by users within Kadi4Mat and users from the same research unit or community can form groups.
The data exchange process happens among users and/or groups through proper access channels taking into consideration the \ac{fair} principles.

\begin{figure}[ht]\centering
	\begin{tikzpicture}[spy using outlines={rectangle,black,magnification=21,size=0.45\textwidth, connect spies}]
		\begin{scope}[spy using outlines={circle,black,magnification=4,size=0.25\textwidth, connect spies}]
			\node (collection) {\includegraphics[width=0.25\textwidth,trim={0 0 0 0},clip]{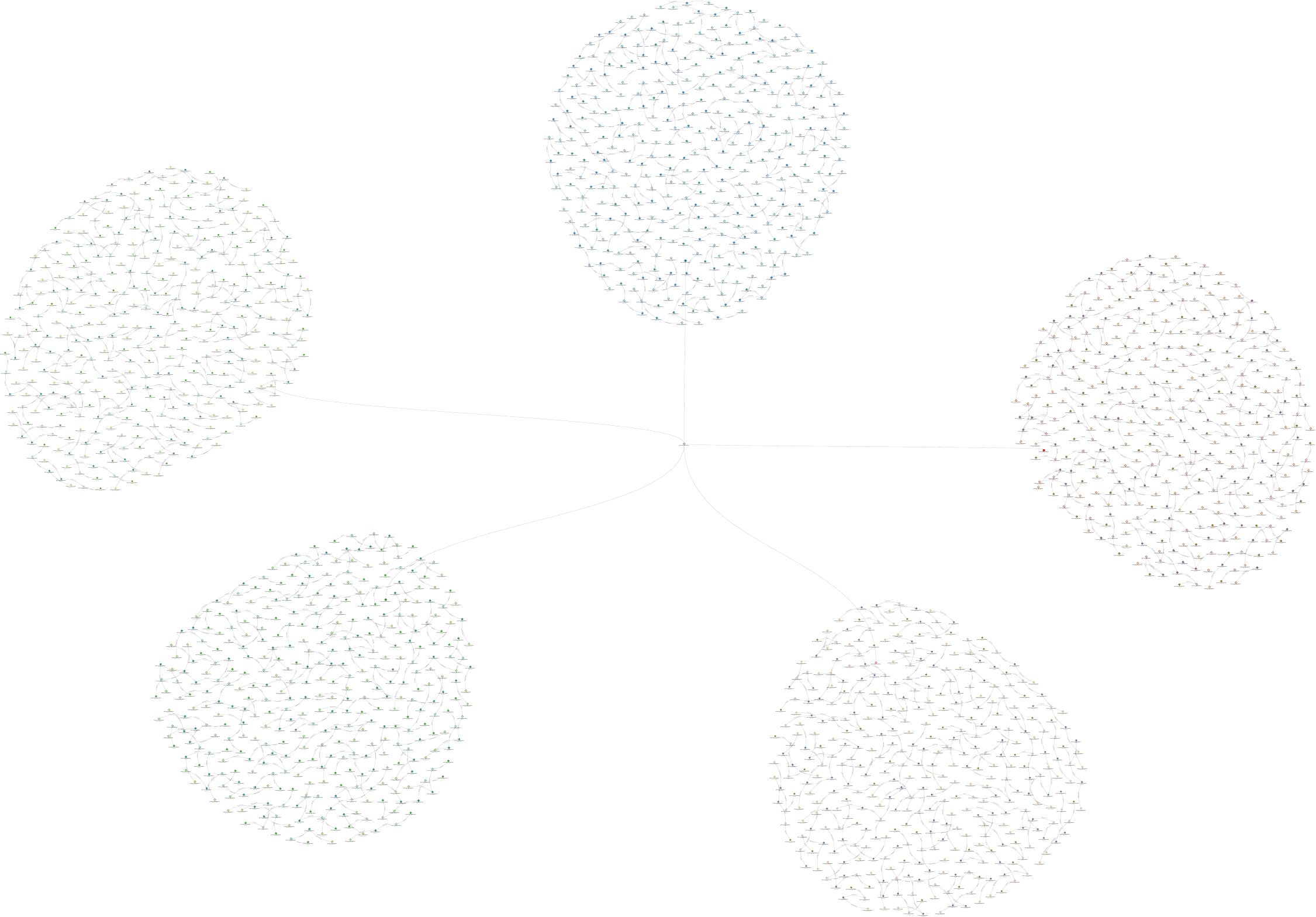}};
			\spy on (1.40,0.1) in node (childcollection)[right] at (1.9,0.25);
		\end{scope}
		\spy on (2.31,-0.09) in node [right] at (5.8,0.25);
	\end{tikzpicture}
	\caption{Visualization of a \kadi{} collection with its records and the links between them, as it is created for each execution of \fe{} pipeline.
	The sample 5 clusters (left) represent 5 Kadi4Mat collections where each collection is a group of records. These collections are children of the main project-level collection, thus appearing like clusters connected to a single point(source). In the middle we see a magnified image of a collection that is a web of inter-linked records, containing all the files that are created in a single pipeline execution.
	To the right, we see records that are part of the magnified collection where the red hexagon symbolizes the collection and the red circles around the records indicate their association to the collection. The inter-linked(related) records shown belong to one specific benchmarking job.
	}\label{fig:collection}
\end{figure}

In this work, for every file a record is created and stored in a collection similar to \cite{tribological} for the execution of the pipeline (c.f. the clusters in ~\autoref{fig:collection}).
Additionally, the records are connected with linked names (c.f. the right part of~\autoref{fig:collection}),
so that it is clear which pipeline execution they belong to and how they are relate to each other.

\subsection{Visualization}\label{subsec:vis}

The collected performance data are visualized in \gf~\cite{grafanaGrafanaQuery} dashboards.
\gf{} is an open source platform for monitoring and observing.
It allows querying and visualizing data in dashboards and panels independently of where it is stored.
\gf{} supports query data from all the standard databases.
Especially, \gf{} plays well with \ac{tsdb}s as it is well suited to visualize time-based data.
With \gf{}, one can create dynamic and interactive plots where the x-axis is the time.

\begin{figure}
	\begin{center}
		\includegraphics[width=0.95\textwidth]{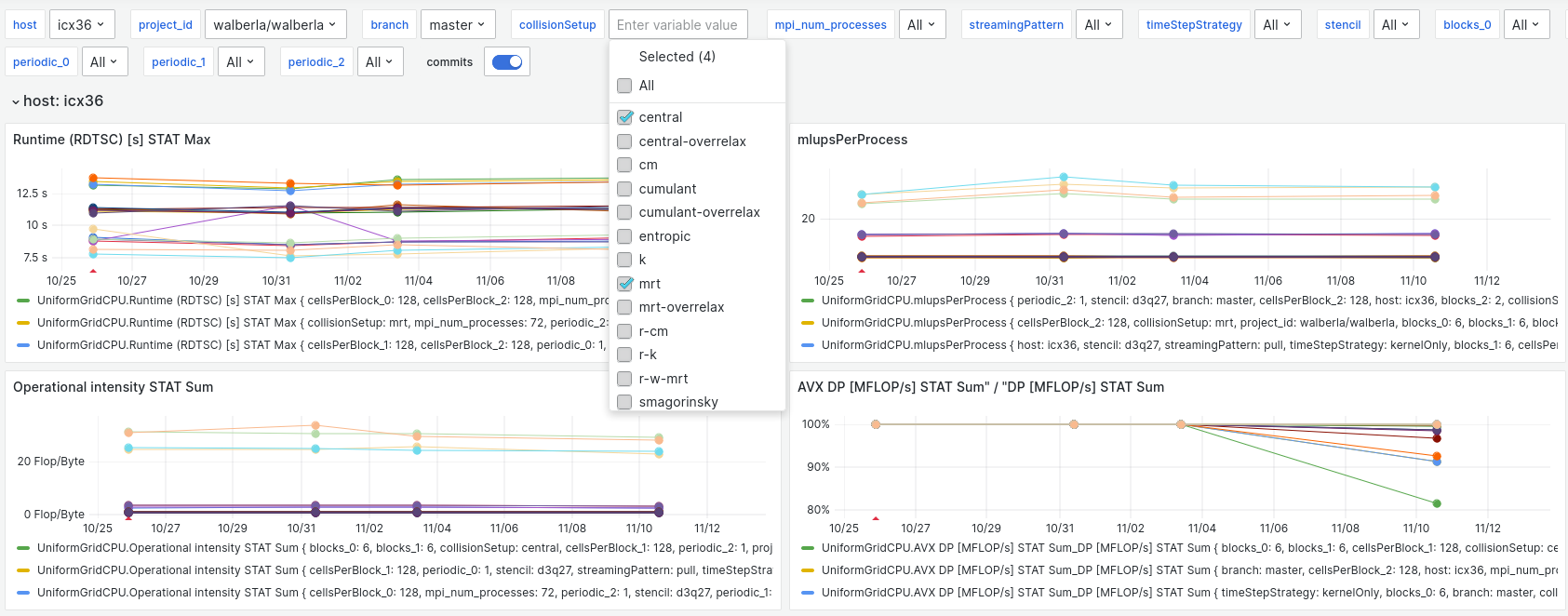}
	\end{center}
	\caption{Visualization Dashboard for the \ac{lbm} Benchmark results with panels and for runtime,
		\ac{mlups} per process, operational intensity and the ratio between vectorized and the total \ac{flop} count.
		The results can be filtered by different simulation parameters, in that case the menu for the collision operator is shown.
	}\label{fig:dash_filter}
\end{figure}

An example of such a dashboard can be seen in~\autoref{fig:dash_filter}.
Each performance metric collected during the benchmark is visualized in a panel.
In ~\autoref{fig:dash_filter}, the panels for the runtime or the \ac{mlups} per process for an \ac{lbm} benchmarks are shown.
The data from the \idb{} is queried and grouped by the different parameter values to connect data points with the same parameter values.
With filters for each run parameter, the results can be adapted interactively.
For example, in the \ac{lbm} benchmarks, it can be observed that the used collision operator influences the performance heavily.
In the dashboard for this benchmark case, there is a filter for the collision operators (c.f. the collisonSetup menu in \autoref{fig:dash_filter}).
This allows the developer to directly compare the results for a chosen set of collision operators.
Similarly, the dashboard for \fe{} has a filter to display only the results for one or more selected solver options.

\begin{figure}
	\begin{center}
		\includegraphics[width=0.85\textwidth,trim={0 0 30cm 0},clip]{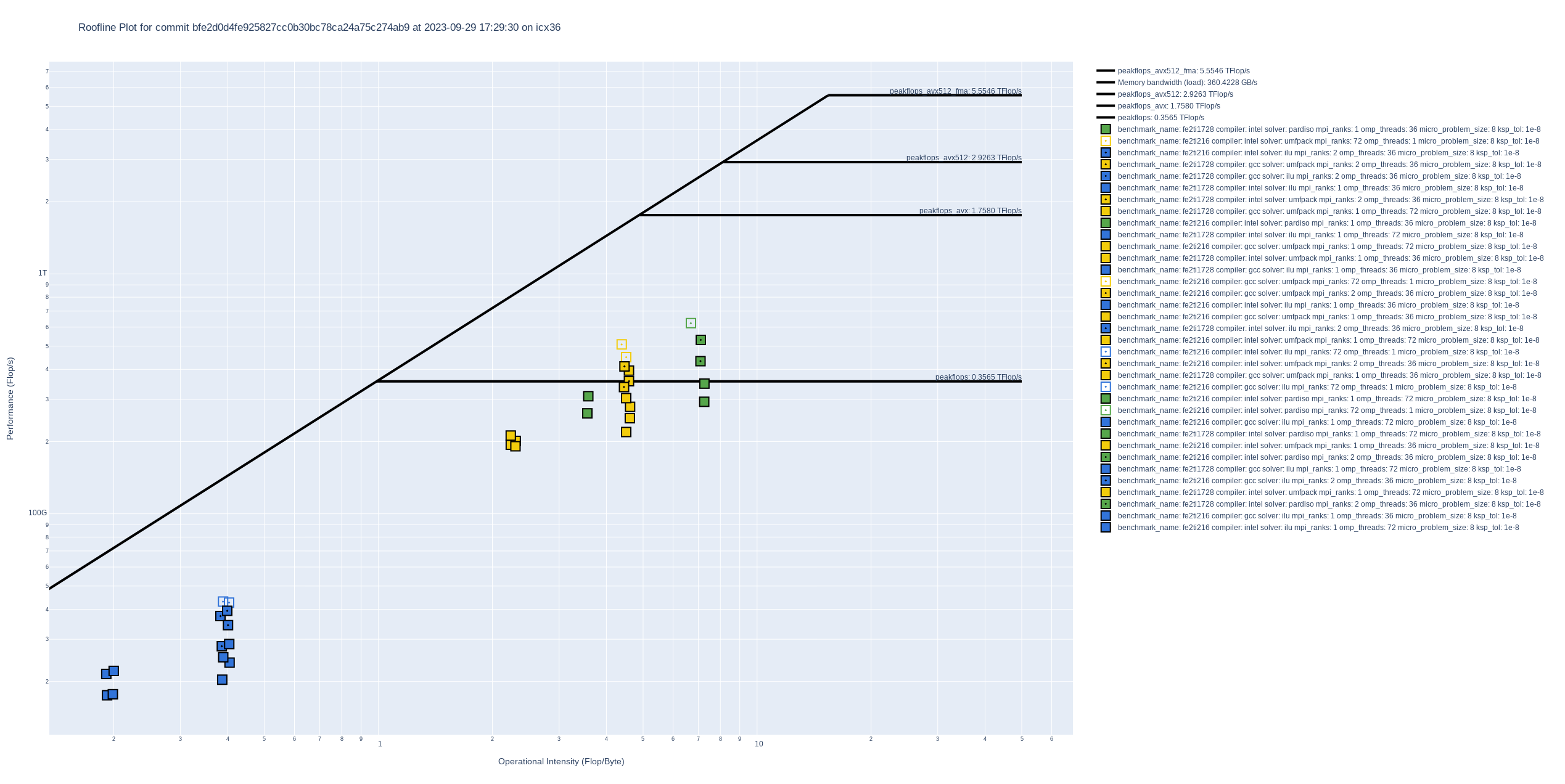}
	\end{center}
	\caption{Screenshot of the roofline plot generated for a \fe{} pipeline execution on a \texttt{icx36} compute node in the Testcluster.
		The green markings stand for the various configurations with the PARDISO solver, the yellow ones for the configurations with the UMFPACK solver and the blue ones use the ILU solver.
	}\label{fig:roofline}
\end{figure}

To track not only the computational performance but also how well the current implementation utilizes the resources of the hardware,
we set the measured result in relation to the theoretical peak performance of the hardware used.
This assessment is based on the roofline model~\cite{williams2009roofline} and
assumes that an application is either limited by the compute resources or by the memory bandwidth.
To get the most realistic upper bounds for the performance, we measured the memory bandwidth and the peak performance for all the compute nodes used in our pipeline with \textit{likwid-bench}.
As benchmark type, we used the different variants of the peak flops, stream, copy, and the load benchmark implemented by \textit{likwid-bench} and stored the results in the \idb{}.
To visualize that, we have, on the one hand, a specific dashboard (c.f.~\autoref{fig:relperf}) and
a \textit{plotly}~\cite{plotly} based python script that generates typical roofline plots (c.f.~\autoref{fig:roofline}).
These roofline plots are generated as interactive HTML files, and can be directly viewed in the browser.

\subsection{Implementation}

The way an application is built and executed is highly project-specific.
Especially if they are as different as our two example applications (~\autoref{tab:comp}).
This results then in different \ac{cb} pipelines.
At the core, such a pipeline needs to be set up and configured individually for each project.
The implementation approach was to build a library that contains all the scripts that can be reused
and then use this to set up a pipeline for each project.
This library now contains some scripts for uploading data,
parsing the output of the different applications, and using profiling tools
and scripts to create the plots.\fl{https://i10git.cs.fau.de/ob28imeq/cb-util}
In addition, the dashboards are specified programmatically with the help of the \textit{grafanalib}~\cite{githubGitHubWeaveworksgrafanalib} python library.
A sketch of how the different components of the pipelines work together can be seen in~\autoref{fig:impl}.
The dashboard for the \fe{} software can be found there\fl{https://www10.cs.fau.de/grafana/d/a3583809-6009-457c-b4d1-71dad2c5b230/fe2ti-benchmarks?orgId=1&refresh=10s},
the dashboard for \wlb{} can be found here\fl{https://www10.cs.fau.de/grafana/d/7HOrefnVk/overview?orgId=1&refresh=10s}.

\subsubsection{Pipeline for \fe{}}

The source code for the \fe{} software, as it is described in \autoref{subsec:feimpl}
is managed in a \gl{} repository that has direct access to the custom \ac{hpc} \gl{} runner (c.f. \autoref{subsec:env}).
So whenever code modifications are pushed to that repository, the pipeline is triggered.
The goal of the pipeline is to automate the benchmarking process using different direct and iterative solvers, compilers, and hardware architectures.
In other words, the pipeline automates the exploration of the parameter space for different execution parameters of the software.

The \textit{fe2ti216} benchmark case, introduced in~\autoref{subsec:feimpl}, use a small macroscopic problem size, so executing the on a single node is suitable.
The benchmark \textit{fe2ti1728} has a macroscopic domain size that is eight times larger.
However, this does not provide any additional valuable insight and would take much longer.
Therefore, it is used in a different execution mode, which emulates the work load of single compute node in a large-scale run.
For that purpose, a benchmark mode was implemented in \fe{}, which omits the macroscopic solve phase and solves only a selection of the \acp{rve}.
To make it possible to perform multiple pseudo time steps, the macro solution is precomputed in a large scale run and finally, in the benchmark run, only read from a file.
The \textit{fe2ti1728} benchmark is used to test the single node performance during the micro solve phase, by solving only 216 of the 1728 \acp{rve}; also compare for the green part in \autoref{fig:fe2_visu}.

Currently, the pipeline is executed on three different compute nodes: \texttt{skylakesp2}, \texttt{icx36} and \texttt{rome1} (c.f.~\autoref{tab:server-configurations}).
For each node, the benchmarks are compiled with the \textit{GCC} compiler, and when possible, the \textit{Intel} compiler is also used.
Also, the different parallelization modes are tested.
The \textit{fe2ti216} benchmark is also executed three times.
First, use only MPI ranks; second, use only OpenMP Threads; and third, use two MPI ranks per node in hybrid mode.
In the current version of \fe{} and the benchmark mode, running the \textit{fe2ti1728} setup in a pure MPI mode is impossible, as this would lead to unequal loads for the different MPI ranks.
The \fe{} pipeline also includes a numerical verification.
Therefore, the solution is compared against a reference solution, and the difference is also visualized in a specific panel.
This gives quick feedback if the parameters' variation also influences the result's numerical quality.
Ultimately, each time the pipeline is invoked, more than 80 different benchmark jobs are generated, which would be tedious to execute manually.

After all benchmark jobs are completed for a pipeline, the script plotting mentioned above automatically creates a roofline plot.
For each run, this script plots the values for operational intensity and \ac{flop} rate measured with \textit{likwid-perfctr} in these roofline plots.
So, in the end, for each instance of the pipeline, we get a roofline plot with all different entries, as shown in~\autoref{fig:roofline}.

\subsubsection{Pipeline for \wlb{}}\label{ssec:pipeline_wlb}

\wlb{} is conceptually a framework for creating \acl{cfd} applications for various scenarios.
Thus, different users use it in production on different hardware architectures.
Therefore, one aspect of the pipeline is to track how the performance on different hardware architectures evolves.
This is done by dynamically generating the benchmark jobs for every supported node in the Testcluster (c.f.~\autoref{tab:server-configurations}), which yields a broad spectrum of architectures.

For \wlb{} already existed a public \gl{} repository\fl{https://i10git.cs.fau.de/walberla/walberla} which the developers actively use.
That repository already uses the \gl{} \acl{ci} features for functional testing.
So, to get access to the \ac{hpc} runner, we created a proxy repository in the \gl{} instance with access to the runner and implemented the \ac{cb} pipeline there.
This pipeline pulls the source code from the actual \wlb{} repository, compiles it for each available node in the Testcluster, and executes the specified benchmarks.
This pipeline is always triggered via the \gl{} trigger API when a commit is made to the default branch of the original \wlb{} repository.
Since the development of \wlb{} is more distributed, and some developers work on individual forks or branches, triggering the \ac{cb} pipeline for these repositories and branches is also possible.
However, this needs to be done manually and can only be done by trusted developers with access to the credentials of the proxy repository.
For the \wlb{} dashboards, an additional filter for repository instance and branch was added.
So that every developer can track the performance impacts of the code changes individually.

As \wlb{} is built from various components, the performance of an application using \wlb{} depends on the used module.
Currently, the pipeline for \wlb{} uses the \textit{UnformGrid\{C,G\}PU} benchmark and the \textit{GravityWaveFSLBM} benchmark~\autoref{subsec:wlb} (c.f.~\autoref{sec:walberla}).
However, it is designed so that new benchmark cases can be easily added in the future.

\begin{figure}
	\begin{center}
		\includegraphics[width=0.95\textwidth]{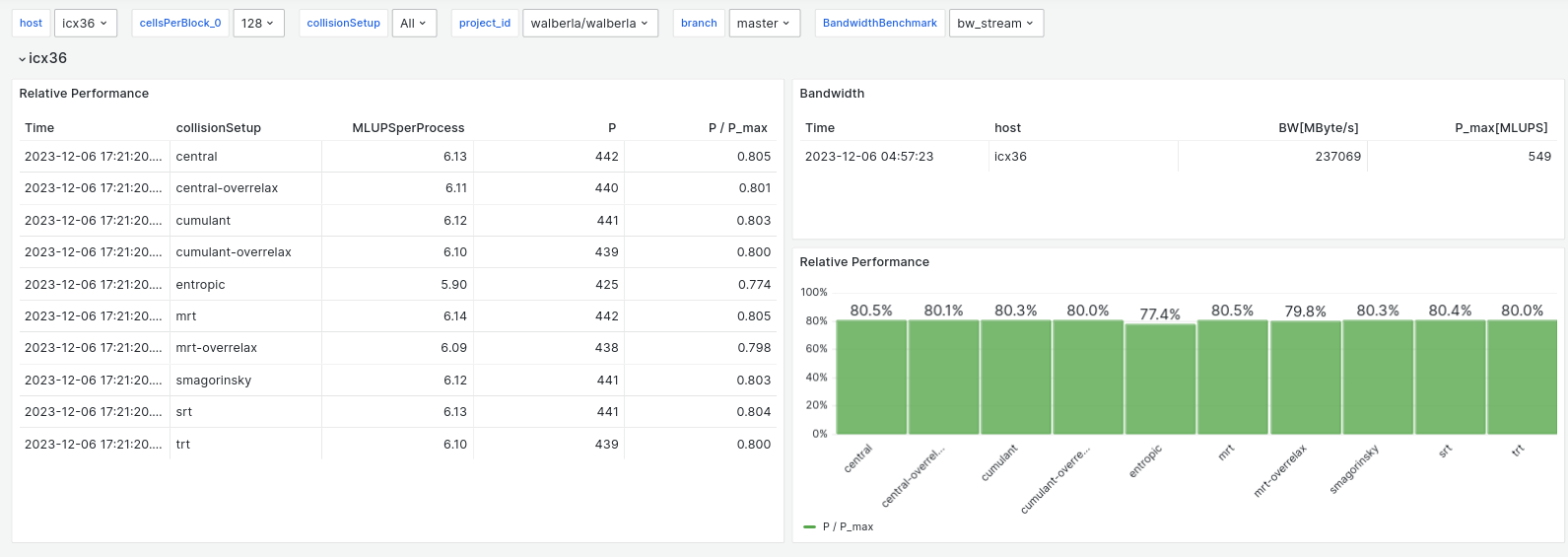}
	\end{center}
	\caption{\gf{} dashboard that shows the achieved performance for different collisions operators in relation to the theoretical peak performance for the uniform grid benchmark on the \textit{Intel Icelake} node.}\label{fig:relperf}
\end{figure}

For the \textit{UniformGridCPU} benchmark, we use the dashboard described in~\autoref{subsec:vis} for the roofline analysis.
In that, we use the procedure that is described in~\cite{doi:10.1177/10943420211016525}, and as Holzer et al., we also assume that the \ac{lbm} application are limited by the memory bandwidth.
This is reasonable for most current hardware architectures, especially those we utilize in our \ac{cb} pipeline.
Therefore, we can calculate the maximum performance $P_{max}$ in \ac{mlups} by dividing the maximal memory bandwidth by the number of bytes read and written during a lattice update.
As the maximum memory bandwidth, we use the results gathered via \textit{likwid-bench} (c.f.~\autoref{subsec:vis}).
In~\autoref{fig:relperf}, the \gf{} dashboard that visualizes the relative performance for the \textit{UniformGridCPU} benchmark is shown.
The \gf{} dashboard in~\autoref{fig:relperf} shows each node the latest benchmark results and calculates the maximum performance for the chosen node.
The user can choose between the different measured bandwidths, which are then used to calculate the maximum performance.

As the \ac{fslbm} is a more complex algorithm with several different steps, it is interesting to track the runtime and the duration of the individual steps.
For this reason, a panel in the dashboard for \textit{GravityWaveFSLBM} benchmark visualizes the different shares of computation, synchronization, and communication in the total runtime (c.f. ~\autoref{fig:fslbmgravitywave}).
For a more detailed breakdown, each group has a panel showing the shares for every sub-step.
This information can be used to identify the most expensive part, which is a good starting point for optimization.

\begin{table}[h]
	\caption{Different benchmark cases currently included in the \acl{cb} pipeline}\label{tab:cases}
	\centering
	\begin{tabular}{ll}
		\toprule
		\multicolumn{2}{c}{\textbf{ \fe{} Benchmark Cases }}                                                        \\
		\midrule
		\multirow{2}{*}{ \textit{fe2ti216} }              & Deformation of dual phase steel with 216 \acp{rve}      \\
		                                                  & with different solvers and parallelization schemes      \\
		\multirow{2}{*}{ \textit{fe2ti1728} }             & same as \textit{fe2ti216} but with 1728 \acp{rve},      \\
		                                                  & but only 216 are solved                                 \\
		\midrule
		\multicolumn{2}{c}{\textbf{ \wlb{} Benchmark Cases }}                                                       \\
		\midrule
		\multirow{2}{*}{\textit{UniformGrid\{CPU, GPU\}}} & Pure \ac{lbm} on a uniform grid, with D3Q27 stencil and \\
		                                                  & different collision operators                           \\
		\textit{GravityWaveFSLBM}                         & Gravity Wave solved with FSLBM                          \\
		\bottomrule
	\end{tabular}
\end{table}

\section{Evaluation}\label{sec:eval}

\subsection{\fe{} Performance Findings}

\begin{figure}
	\begin{center}
		\includegraphics[width=0.95\textwidth, trim={0 0 0 0},clip]{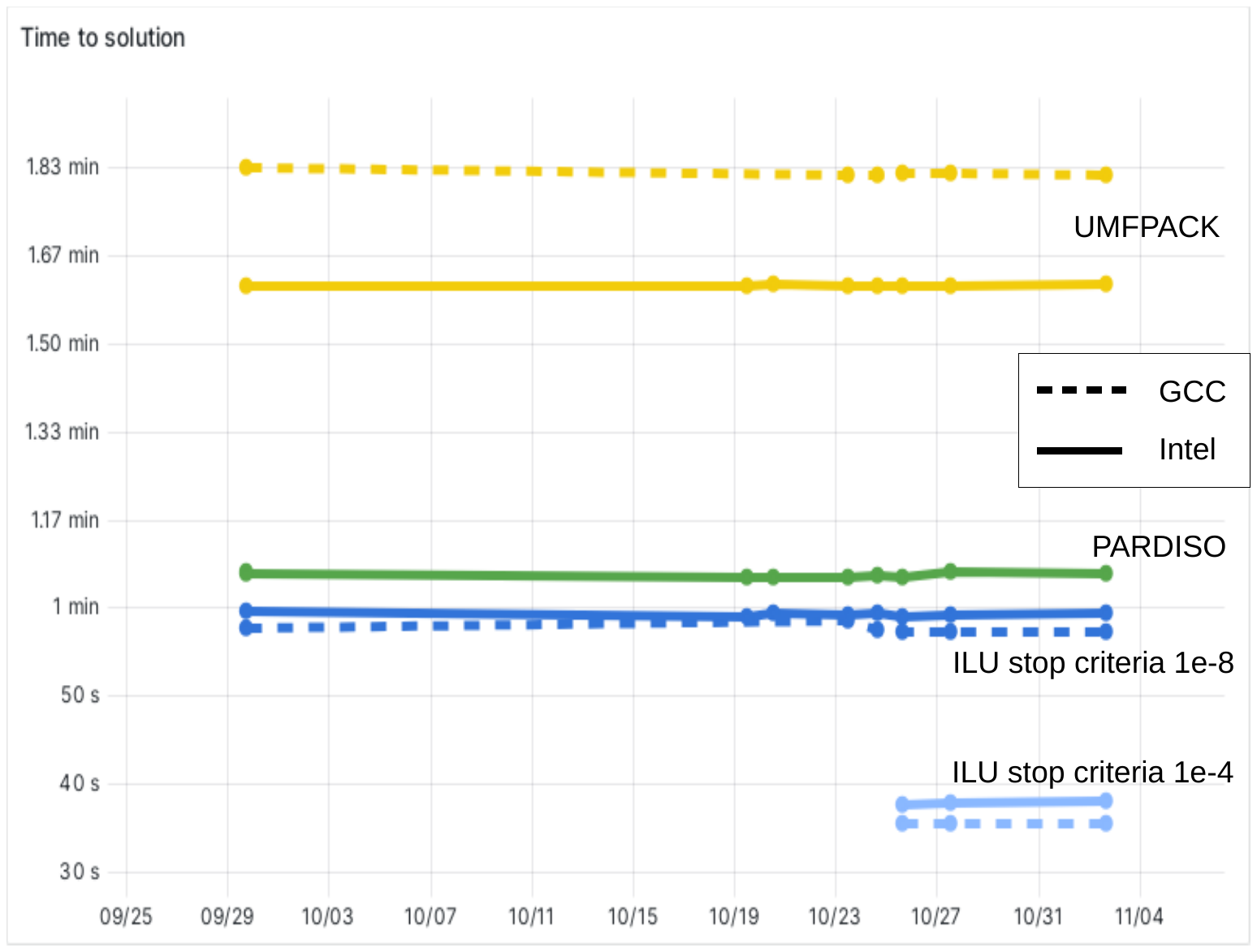}
	\end{center}
	\caption{\ac{tts} for the \textit{fe2ti216} case of all solvers and on the \texttt{icx36} node using 72 MPI ranks.
		The color encodes the solver, where green is PARDISO, yellow UMFPACK and blue ILU.
		For ILU the dark blue variant uses 1e-8 as stopping tolerance and the light blue variant uses 1e-4 as stopping tolerance.
		Dashed lines are compiled with gcc and use OpenMPI and solid lines are compiled with the intel compiler and use IntelMPI.}\label{fig:ttsall}
\end{figure}

\begin{figure}
	\begin{subfigure}[t]{0.5\textwidth}
		\includegraphics[width=0.95\textwidth]{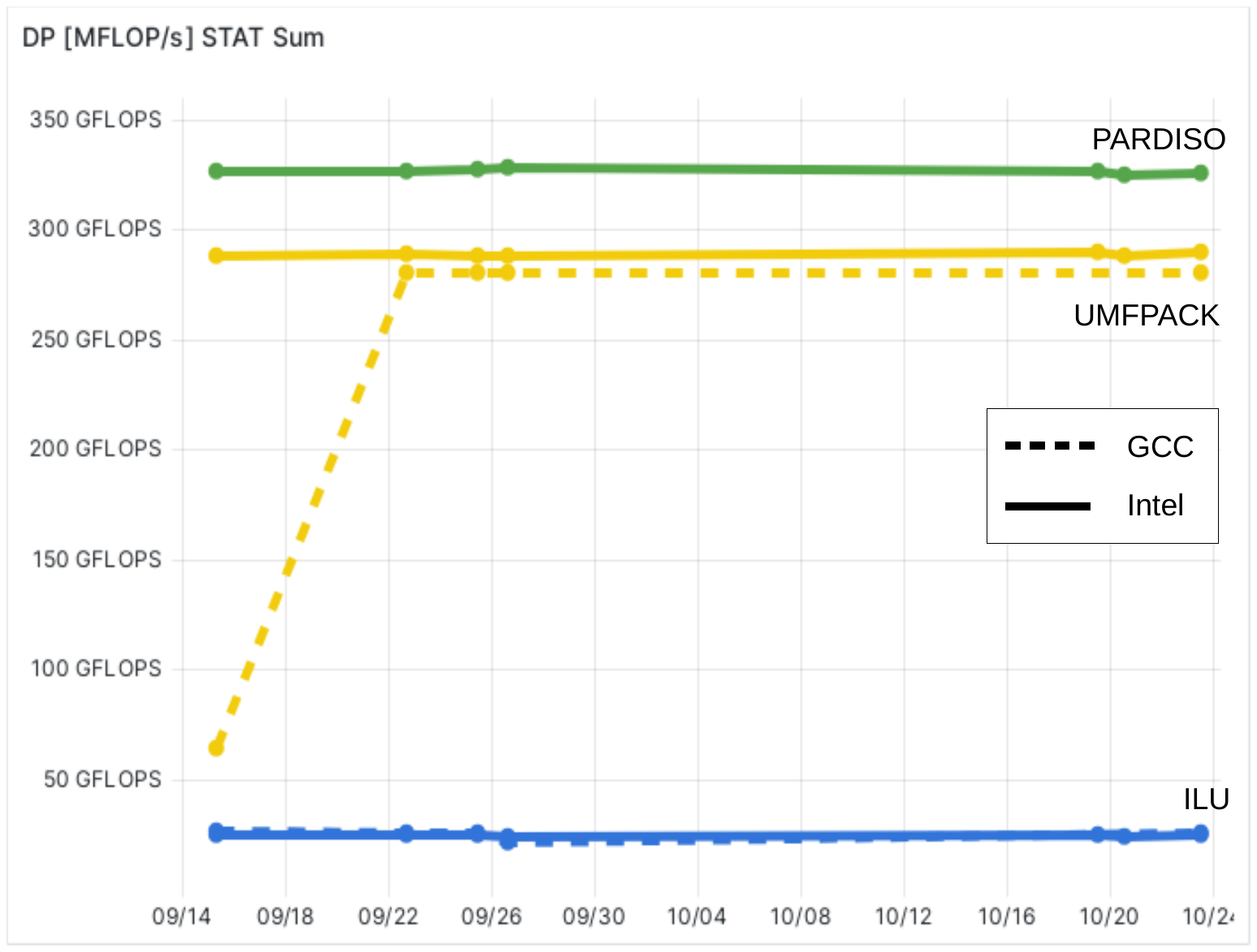}
		\caption{Performance in \ac{flops}}\label{fig:perfmpi}
	\end{subfigure}
	\begin{subfigure}[t]{0.5\textwidth}
		\includegraphics[width=0.95\textwidth]{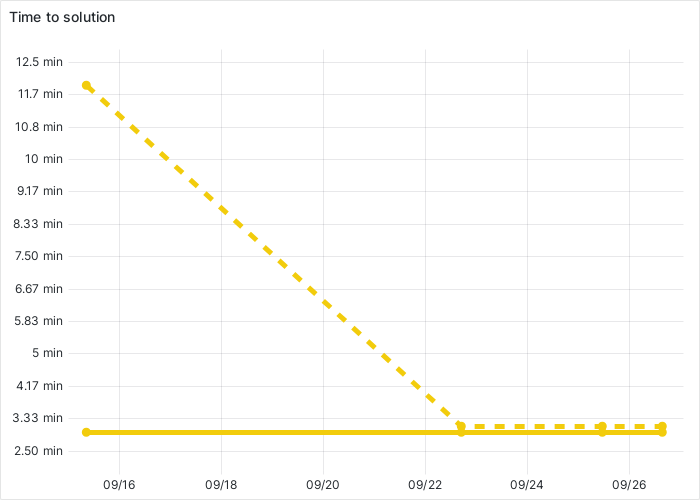}
		\caption{\ac{tts} for the UMFPACK solver.}\label{fig:umfpack}
	\end{subfigure}
	\caption{Results for the \textit{fe2ti216} benchmark on the \texttt{skylakesp2} node using pure MPI parallelization.
	The colors encode the solver where green stands for PARDISO, yellow for UMFPACK and blue for ILU.
	Dash lines are compiled with gcc, solid lines use the Intel compiler.}
\end{figure}

When comparing the \ac{tts} of the different setups, it becomes evident that the setups using ILU are the fastest.
Especially when the stopping tolerance for the Krylov subspace solver is set to a higher value.
All data points shown in~\autoref{fig:ttsall} are created with the same version of the software, and over the different runs, the results remain stable.
\autoref{fig:ttsall} shows the different \ac{tts} on the \texttt{icx36} node with a pure MPI parallelization, but for all other nodes and parallelization schemes, the picture looks similar.
Close to the iterative solver ILU is the PARDISO solver, and the slowest one is UMFPACK complied with \textit{gcc}.
When relaxing the stopping criteria for the iterative solver, the \ac{tts} becomes even lower.
In contrast to the direct solvers, the iterative solver is doing less work, as seen in~\autoref{fig:perfmpi}.
Comparing the performance in \ac{flops}, the PARDISO solver reaches the highest value, and ILU only achieves around 25 G\ac{flops} (compare ~\autoref{fig:perfmpi}).
Since Newton's method for the macroscopic problem still converges, the inexact solution for the microscopic problem is sufficiently exact and it is not necessary to use a higher accuracy here.

In~\autoref{fig:perfmpi}, there is a jump in the performance for the setup with UMFPACK where PETSc is compiled with \textit{gcc}.
Also in~\autoref{fig:umfpack}, where the \ac{tts} of the UMFPACK setups is shown, it can be observed that the \textit{gcc} version
has huge decrease in the \ac{tts} at the same time.
Also, for all the other setups, we observed a similar jump in performance for the UMFPACK solver between the two compilers.
The reason, therefore, was that the PETSc setup compiled with the \textit{Intel compiler} was linked against the MKL routines,
whereas the \textit{gcc} version was linked against the slower reference routines provided by PETSc.
It was possible to close that gap by compiling PETSc against the BLIS routines~\cite{BLIS1}.

\begin{figure}
	\begin{subfigure}{0.5\textwidth}
		\includegraphics[width=0.95\textwidth]{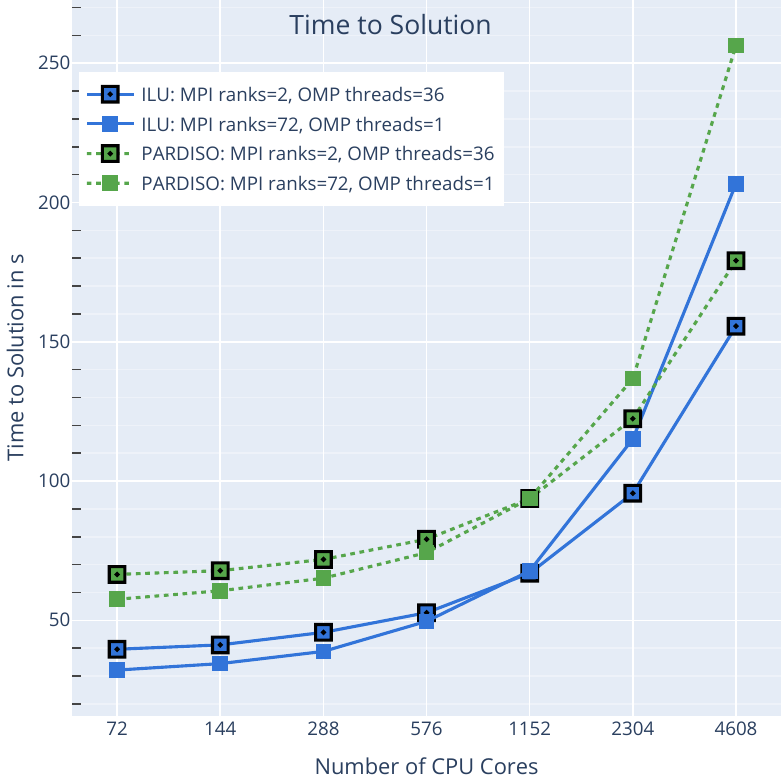}
	\end{subfigure}\label{fig:weaktts}
	\begin{subfigure}{0.5\textwidth}
		\includegraphics[width=0.95\textwidth]{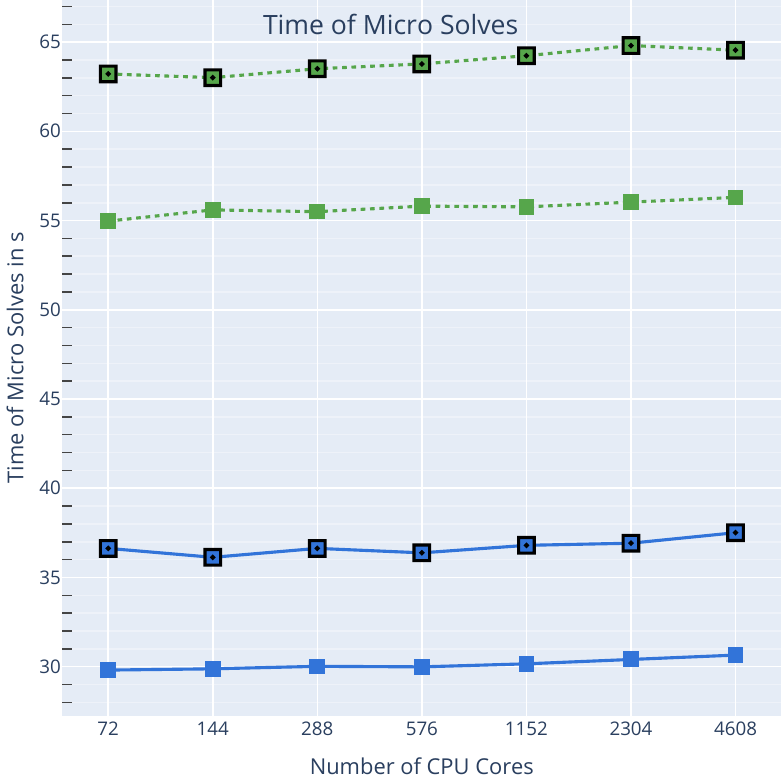}
	\end{subfigure}\label{fig:weakmicro}
	\caption{Weak Scaling on the Fritz supercomputer, with 216 \ac{rve}s per node from 1 to 64 nodes.
	The blue solids lines stand for the iterative solver (ILU) and green for the PARDISO solver.
	The squares with dot stand for hybrid parallelization and the squares without use pure MPI parallelization.
	Let us note that we sum up the microscopic solution times over all Newton iterations in all load steps.
	}\label{fig:weak}
\end{figure}

To verify that the results gathered via the \ac{cb} pipeline yield meaningful insights for runs at a large scale, we conducted a weak scaling benchmark on the Fritz supercomputer at \nhr{}~\cite{fauFritzNHRFAU}.
The compute nodes there also contain \icx{}\textit{s} with 72 cores, which are also used in the \ac{cb} setup.
For our comparison we selected the two fastest solvers from the \ac{cb} pipeline: the iterative Krylov subspace solver with the relaxed stopping criteria and the PARDISO solver.
We tested each solver once in a pure MPI setup and once in an MPI/OpenMP parallelization setup.
We use a setup inspired by the \textit{fe2ti216} benchmark, with exactly 216 \acp{rve} per node, and scale it from 1 to 64 nodes.
In~\autoref{fig:weak}, we can see that for the single node run, the results are nearly identical to the \ac{cb} pipeline results.
The iterative solver needs around 40 seconds as \ac{tts}, and the PARDISO setup needs around 60 seconds~\autoref{fig:ttsall}.
The results on Fritz are slightly better compared with the results on the Testcluster.
The simple explanation is that in the large-scale runs, the CPU frequency was not fixed to 2.0 GHz, which is always done in the \ac{cb} pipeline.
For the single node runs, we can also see that the \ac{tts} is entirely dominated by the time that is used for solving the \acp{rve}.
This is due to the very small macroscopic problem size resulting in a negligible solution time in the sparse direct solver package on the macro scale.
If we look at the time used for solving the micro problems, it can be seen that the time remains almost constant for all nodes and different solvers.
This shows that we can gather meaningful results about the performance of the microscopic solving phase with our \ac{cb} on a single node,
since the microscopic phase scales nearly ideally and thus the \ac{tts} is nearly constant on one or multiple nodes.
Another interesting finding is that the time for micro-solving is slightly shorter if the application uses only MPI for parallelization.
For ILU, it is around 6 seconds, and for PARDISO, it is around 8 seconds for all tested numbers of nodes.
This behavior can also be seen in the \ac{cb} setup and on the other hardware architectures there and might be an overhead introduced by the OpenMP runtime.
Further, in the \ac{cb}, we see slightly higher data volume transferred during these hybrid jobs.
Here, further investigations are necessary and caused by the \ac{cb}, which again shows the benefit of this approach.

While the RVE solves scale more or less perfectly to more nodes, the overall \ac{tts} does not.
This is expected since, with increasing macroscopic problem size, the impact of solving the macroscopic problem with a sequential sparse direct solver becomes more and more dominant.

\begin{figure}
	\begin{center}
		\includegraphics[width=0.6\textwidth]{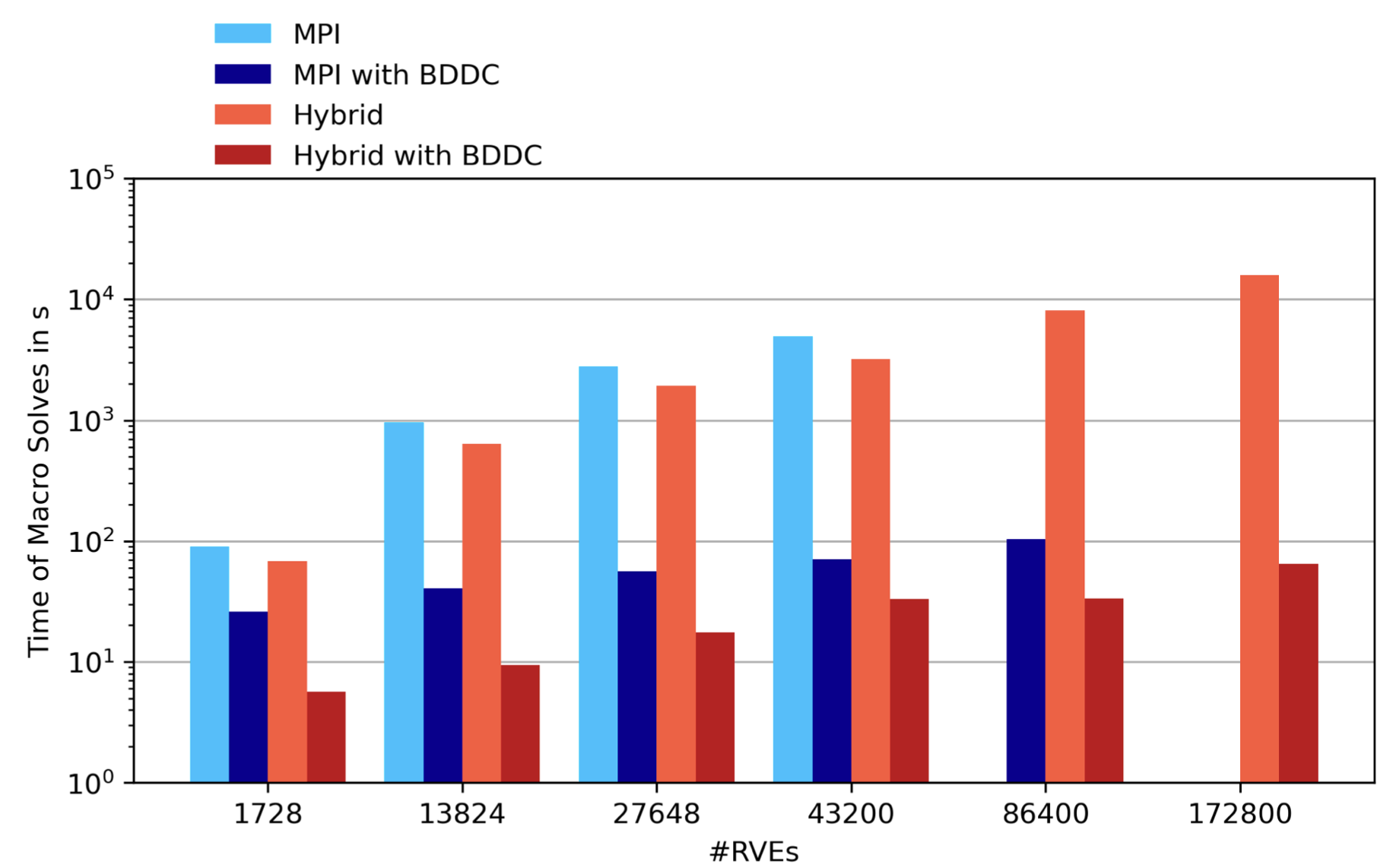}
	\end{center}
	\caption{Comparison of the weak scaling of two different macroscopic solvers (sequential MKL-Pardiso and parallel BDDC) on the JUWELS supercomputer using 9 to 900 nodes with 48 cores each.
	On each node 192 RVEs are handled. The hybrid configuration uses 2 MPI ranks per node and 24 threads per rank. The time for all macroscopic solves in all Newton steps is summed up.
	}\label{fig:weak_juwels}
\end{figure}

This scaling bottleneck can be overcome using a parallel solver for the macroscopic problem instead.
In FE2TI, there is the option to use the parallel \ac{bddc} domain decomposition solver on a subset of MPI ranks, which improves the overall weak scalability.
Since this is not the focus of this article, we only show some additional scaling results obtained on JUWELS~\cite{fzjuelichJUWELS} to prove and explain the scaling behavior; see Figure~\ref{fig:weak_juwels}.

For 1 to 8 nodes, the pure MPI parallelization achieves the lower \ac{tts}; with 16 nodes, they are more or less equal, and for the higher node counts, the hybrid parallelization is better.
This can be explained mainly by the MPI communication overhead.
With a pure MPI, we have as many MPI ranks as we use CPU cores, and each of them needs to communicate with each other during the macro-solve phase.
Using the hybrid parallelization, the number of MPI ranks is only the number of nodes times two, so fewer MPI ranks need to communicate with each other.
Although the pure MPI parallelization seems to perform better for the microscopic RVE problems, the hybrid version outperforms the pure MPI parallelization due to lower communication overhead.

\subsection{\wlb{} Performance Findings}

\begin{figure}
	\begin{center}
		\includegraphics[width=0.95\textwidth]{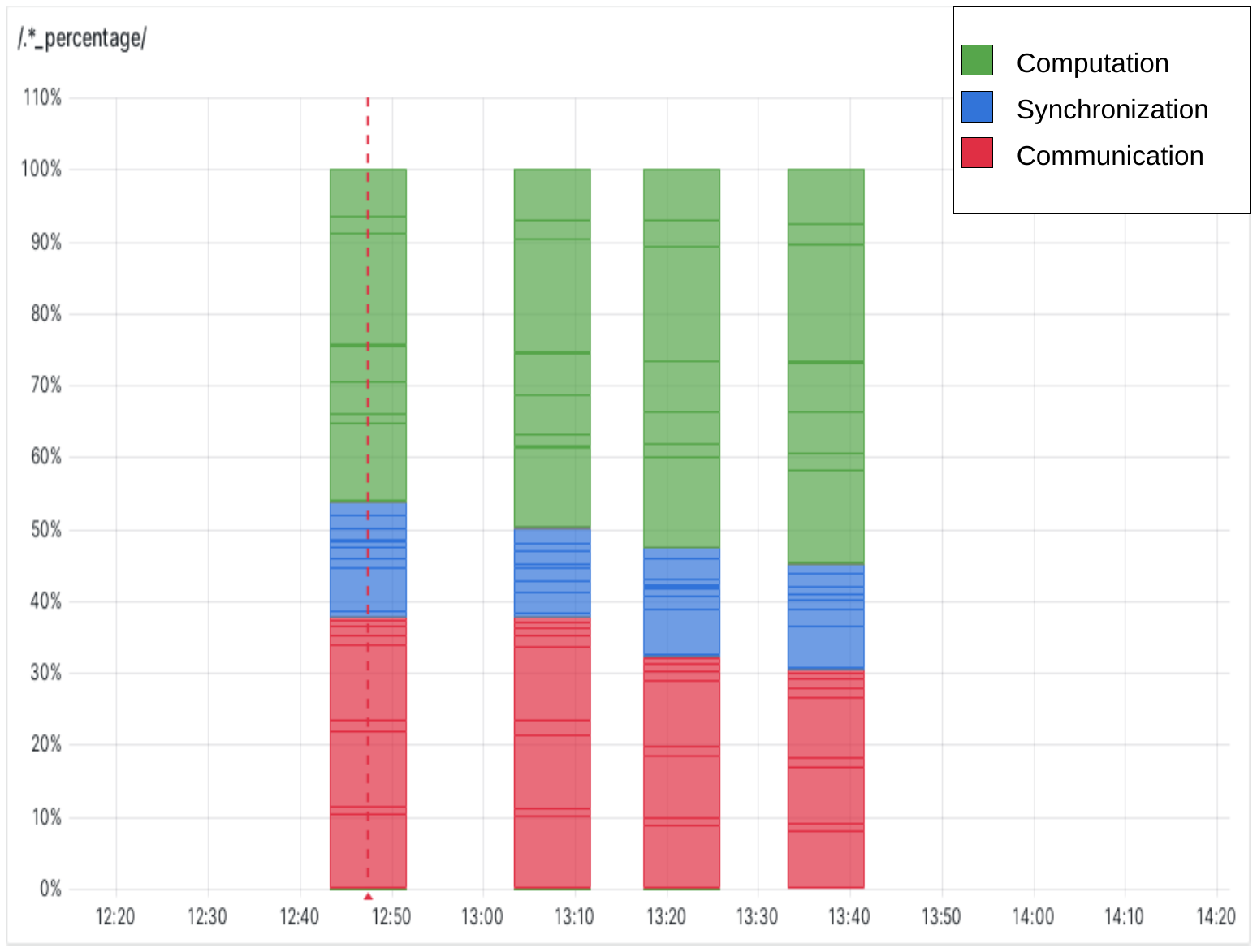}
	\end{center}
	\caption{Grafana panel which shows the distribution of the simulation time for the gravity wave benchmark.
	The red part is for Communication, the blue for synchronization and the green is for computation.
	The left bars are the results for the \texttt{skylakesp2} node, for the \texttt{icx36} node, for the \texttt{rome1} node and for the \texttt{genoa2} node.
	In that order from left to right.}\label{fig:fslbmgravitywave}
\end{figure}

In~\autoref{fig:relperf} we can see that if the maximum performance is based on the stream benchmark (around 237 GB/s on the Icelake node) the \textit{UniformGridCPU} achieves around 80\% of it.
Additionally, we want to get the current performance state regarding the FSLBM implementation in \wlb{} with the \textit{GravityWaveFSLBM} benchmark case.
It is implemented such that after each computation step described in \autoref{subsec:fslbm}, there is synchronization and communication.
In the benchmark reported in this paper, we enforce an artificial synchronization after each computation and before the communication starts.
In this way we can distinguish between the synchronization and communication time of the algorithm.
As described in \autoref{subsec:wlb}, we use a 2D block-decomposition to use an artificially induced perfect load balancing, such that we can accurately analyze the performance of FSLBM, without having further parts influencing the performance.
The benchmark always uses the maximum available cores per node, which differs depending on the architecture.
The domain size is scaled with the number of cores, so every core has one block ($32^3$) with one gravity wave after initialization.
\autoref{fig:fslbmgravitywave} shows the CB-results for different architecture for the gravity wave benchmark test case in distributions of the total simulation time for the different grouped parts of the FSLBM.
Here, the computation is reported with approximately 45-55\% of the total simulation time depending on the architecture, synchronization takes 12-18\%, and communication ranges from 30-38\%.
We have here two notable findings.
First, the communication overhead with 30-38\% is high, which can be explained by choosing a relatively small block size per core with $32^3$ cells.
Second, there should be almost no synchronization overhead because we artificially chose an initialization where all cores have an identical workload.

\begin{figure}
	\centering
	\begin{subfigure}[t]{0.49\textwidth}
		\subcaptionbox{Simulation time\label{fig:t_FSLBM}}{\includegraphics[width=0.95\textwidth]{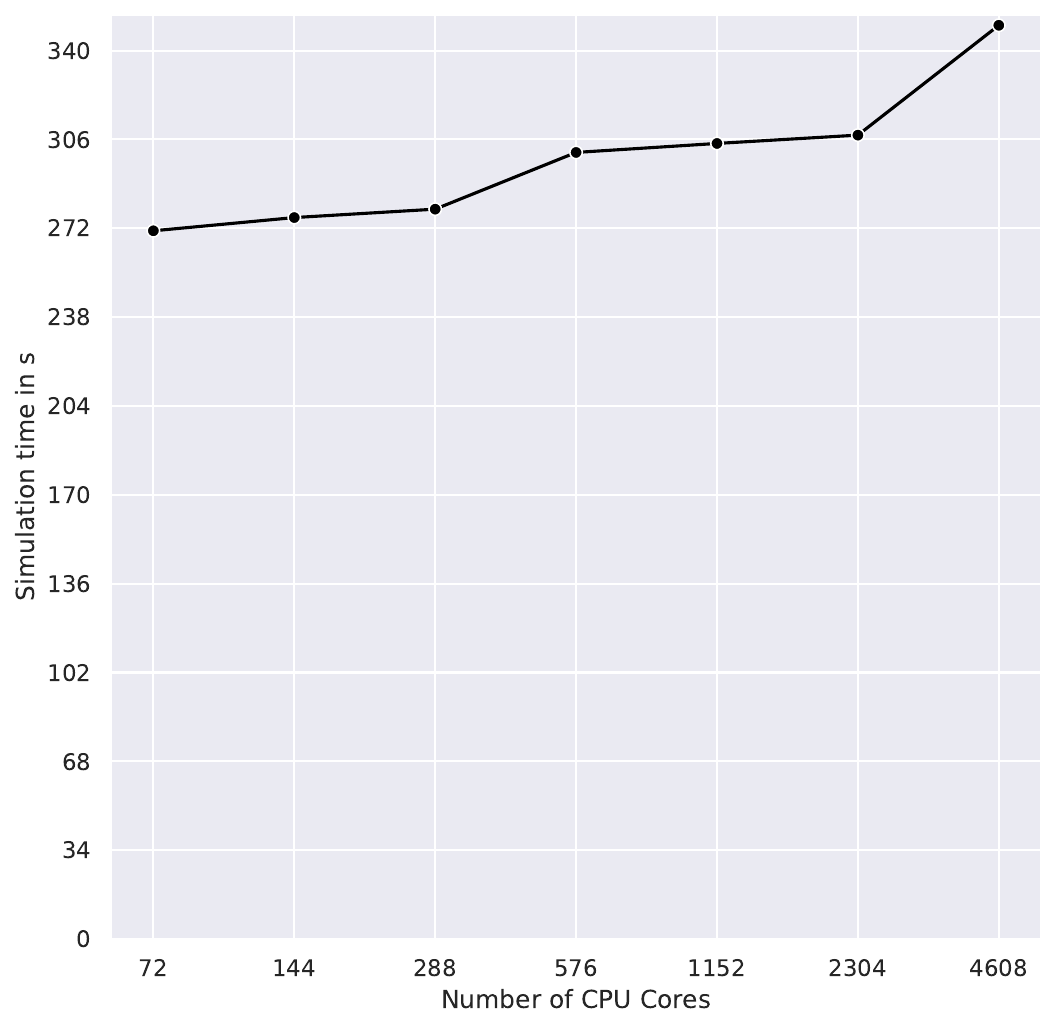}}
	\end{subfigure}
	\begin{subfigure}[t]{0.49\textwidth}
		\subcaptionbox{Execution time of communication, computation and synchronization\label{fig:ttpkc_FSLBM}}{\includegraphics[width=0.95\textwidth]{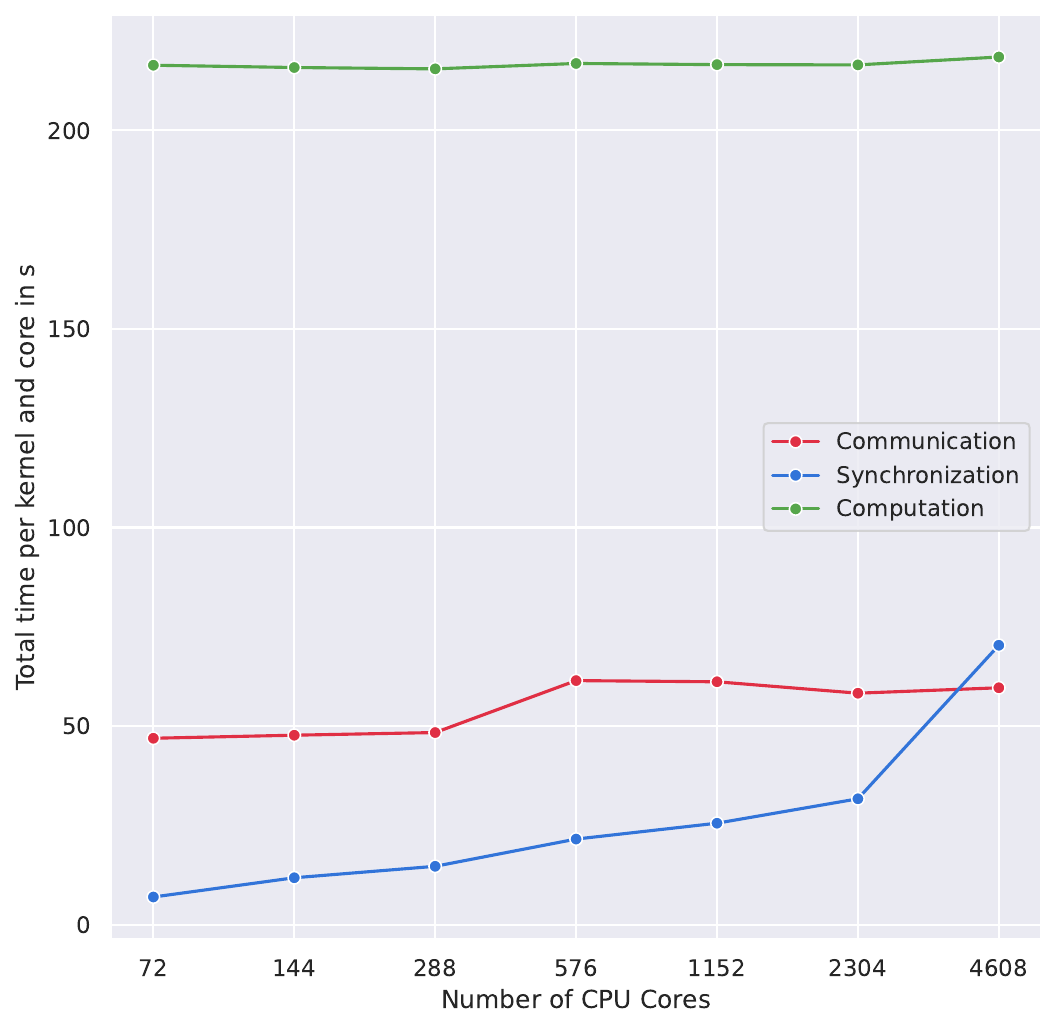}}
	\end{subfigure}
	\caption{Weak scaling results on the Fritz supercomputer from 72 to 4608 CPU cores. Which each node having 72 cores.}\label{fig:perf_FSLBM}
\end{figure}

To further investigate those findings, we run a weak-scaling benchmark on the Fritz supercomputer from 1 to 64 nodes with the same initialization but this time using a block size of $64^3$ cells per core.
For this benchmark, we allocated the 64 nodes once and then ran the benchmarks with a different number of nodes.
This approach ensures that the node topology does not further influence the reported results due to multiple node allocations and benchmark submissions.
The results for the total simulation times for 1 to 64 nodes are shown in~\autoref{fig:t_FSLBM}.
We see a slightly increasing simulation time with increasing cores, with two steps of degradation from 4 to 8 nodes and from 32 to 64 nodes.
Taking a look into the execution time per kernel and per core for the communication, computation, and synchronization part of the FSLBM algorithm, shown in~\autoref{fig:ttpkc_FSLBM}, we can see where these jumps and the overall increase in simulation time is coming from.
The first jump in simulation time between 4 and 8 nodes is due to an increasing overhead for communication and synchronization. In contrast, the second jump is solely caused by the synchronization.
While the increasing time for communication can be explained as a non-optimal allocation of the nodes in the system, gives rise to further investigations that are currently ongoing.
In this way the \ac{cb} demonstrates its usefulness as tool to help develop efficient simulation software in a systematic way.
Furthermore, the results show a perfect scaling for the computation and an almost perfect scaling for the communication.

\section{Related Work}\label{sec:relwork}

In \cite{10.1145/3324989.3325719}, Anzt et al. presented a conceptually similar \ac{cb} setup for their linear algebra library GINGKO.
Their performance evaluation framework uses \gl{} \acl{ci} functionalities to submit benchmarks to \ac{hpc} resources, and the results are visualized via a self-developed web interface.
In \cite{DBLP:journals/corr/abs-1812-03149} the authors present an \idb{}~\cite{influxdataInfluxDBRealtime} and \gf{}~\cite{grafanaGrafanaQuery} based approach for storing and presenting benchmarks results,
with the difference that they are not gathered on \ac{hpc} resources.
A jacamar~\cite{gitlabExascaleComputing} based approach for a hypersonic aerodynamics code is presented in \cite{meloneVerifyingFunctionalityPerformance2023}.
Further, some other frameworks like the Helmholtz Anlytics Toolkit (HEAT)~\cite{heat2020} employs a \ac{cb} pipeline
to avoid performance degredations~\cite{claudia_comito_2023_8060498}.
In \cite{10.1145/3624062.3624135} Pearce et al. proposed the concept of \textit{collaborative continuous benchmarking}
which is currently more focused on evaluating the \ac{hpc} systems themselves.
Therefore, they focus on interoperability between different \ac{hpc} systems, which is achieved by automating the build process with
\textit{spack}~\cite{10.1145/2807591.2807623} and \textit{ramble}~\cite{githubGitHubGoogleCloudPlatformramble} in combination with streamlined evaluation process.

\section{Conclusion}\label{sec:outro}

This paper presented our \emph{\acl{cb}} strategy and our implementation for two different \ac{hpc} codes.
Each of them had its requirements and challenges to be considered during development.
We showed the benefits of \ac{cb} and that it enables developers to a more performance-centric development process
and reveals performance degradation introduced by code changes immediately.
In the long run, storing the results in a \ac{fair} way makes tracking the performance changes over a long period possible.
Furthermore, it makes it possible to trace which code change leads to a change in the performance characteristic.
This is not only interesting when specific performance optimizations are included in the application, but also when the application is extended to simulate a more complex scenario or to resolve phenomena more accurately.
In this case, it is also possible to easily quantify the impact on performance compared to the previous version.
In our interactive visualization, not only are the pure performance metrics shown, but these are also related to the capabilities of each machine.
Thus, the developers get quick feedback on how well the hardware resources are utilized.

For the \fe{} software package, we found out that the fastest solution is to use an inexact solver for solving the micro problems
and showed that this also holds for runs with up to 4608 CPU cores.
Further, this is a positive finding as this solver does not rely on a vendor-specific implementation and can also be used on clusters that use AMD CPUs.
For \wlb{}, we found out that the implementation of different \ac{lbm} variants already utilizes the hardware quite well.
For the \ac{fslbm} implementation, the \ac{cb} pipeline revealed a few bottlenecks and starting points for further investigations.
Regarding \wlb{}, we discovered that the implementation of various \ac{lbm} variants already makes efficient use of the hardware.
However, the CB pipeline identified some bottlenecks and areas for further investigation in the FSLBM implementation.

Future extensions to the \wlb{} pipeline may include support for other hardware architectures, such as AMD GPUs or the upcoming accelerated processing units from AMD or Nvidia.
Additionally, new benchmark cases may be added to test the performance of other framework aspects.
Furthermore, another next step is to add support for multi-node benchmarks and automate weak scaling runs and their evaluation.
It may also be beneficial to include support for the \acl{ci} services at other computing centers, enabling the use of \ac{hpc} resources beyond the local compute center.

\section*{Acknowledgements}
This project has received funding from the Deutsche Forschungsgemeinschaft (DFG, German Research Foundation) under the project Numbers 433735254 and 434946896.
This work has also received funding from the European High Performance Computing Joint Undertaking (JU) and Sweden, Germany, Spain, Greece, and Denmark under grant agreement No 101093393.
The authors gratefully acknowledge the Gauss Centre for Supercomputing e.V. (www.gauss-centre.eu) for funding this project by providing computing time through the John von Neumann Institute for Computing (NIC) on the GCS Supercomputer {\bf JUWELS} at J\"ulich Supercomputing Centre (JSC).
The authors gratefully acknowledge the scientific support and HPC resources provided by the Erlangen National High Performance Computing Center (NHR@FAU) of the Friedrich-Alexander-Universität Erlangen-Nürnberg (FAU) under the NHR project "HEISSRISSE - Massively Parallel Simulation of the Melt Pool Area during Laser Beam Welding using the Lattice Boltzmann Method". NHR funding is provided by federal and Bavarian state authorities. NHR@FAU hardware is partially funded by the German Research Foundation (DFG) – 440719683.

\printbibliography{}

\end{document}